\documentclass[aps,prd,letterpaper,10pt,twocolumn,superscriptaddress]{revtex4-2}
\usepackage[utf8]{inputenc}
\usepackage[T1]{fontenc}
\usepackage[english]{babel}
\usepackage{ragged2e}
\usepackage{amsmath}
\usepackage{graphicx} 
\usepackage{physics} 
\usepackage{braket} 
\usepackage{times}
\usepackage{hyperref}
\hypersetup{colorlinks=true,linkcolor=blue,citecolor=cyan}
\usepackage{float}

\newcommand{\ee}{\mathrm{e}}             
\newcommand{\ii}{\mathrm{i}}             

\newcommand{\Ai}[1]{\text{Ai}\left(#1\right)}

\newcommand{\tbraket}[1]{\langle #1 \rangle_\beta}
\newcommand{\pbraket}[2]{\langle #1 | #2 \rangle}
\newcommand{\ppbraket}[3]{\langle #1 | #2 | #3 \rangle}

\newcommand{\orcid}[1]{\href{https://orcid.org/#1}{\textsuperscript{\,\includegraphics[height=2ex]{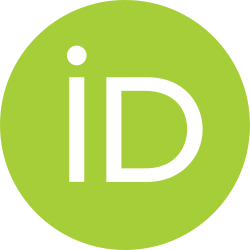}}}}

\begin{document}

\title{Operator Ordering in the Relativistic Quantization: \\ Specific Heat in the Rindler Frame}

\author{Karol Sajnok\orcid{0009-0004-5899-8923}}
\email{ksajnok@cft.edu.pl}
\affiliation{Institute of Theoretical Physics, University of Warsaw, ul. Pasteura 5, 02-093 Warsaw, Poland}
\affiliation{Center for Theoretical Physics, Polish Academy of Sciences, Aleja Lotników 32/46, 02-668 Warsaw, Poland}

\author{Kacper Dębski\orcid{0000-0002-8865-9066}}
\email{k.debski@uw.edu.pl}
\affiliation{Institute of Theoretical Physics, University of Warsaw, ul. Pasteura 5, 02-093 Warsaw, Poland}

%
\date{\today}
\begin{abstract}
We introduce a covariant canonical quantization for a particle in curved spacetime that tracks operator‐ordering ambiguities. Parameterizing spatial and temporal ordering, we derive a Hermitian Hamiltonian with leading quantum‐relativistic corrections. In a uniformly accelerated frame, we show the semiclassical heat‐capacity approximation misses these effects and then develop a perturbative quantum treatment using Airy‐function modes to obtain analytical first‐ and second‐order energy shifts. Including these shifts in the partition function yields nontrivial, ordering‐dependent specific‐heat corrections. Numerical studies for electrons in extreme electric fields and ultra‐light particles in strong gravitational fields demonstrate that these corrections become significant at intermediate temperatures. Enforcing the Tolman–Ehrenfest relation for spatial temperature variation further modulates the heat‐capacity profile. Our results suggest that precision calorimetry in laser‐acceleration or analogue gravity setups could probe quantum‐ordering effects in relativistic regimes.
\end{abstract}

\maketitle
\footnotesize
\fontsize{9pt}{11pt}\selectfont
\section{Introduction}  \label{sec:i}
The quest to unify quantum mechanics and general relativity remains one of theoretical physics’ greatest challenges: how to describe particles at quantum scales while preserving the dynamical structure of spacetime? Numerous frameworks have been proposed, yet none yield fully testable predictions.

Quantum Field Theory (QFT) \cite{peskin2018, feynman2006, feynman2018, zee2010}, while providing a robust relativistic framework for describing quantum fields, does not inherently address the quantum nature of gravity. Instead, it adapts quantum mechanics to the classical structure of general relativity, leaving the quantization of spacetime itself elusive. Geometric quantization \cite{kostant1970, woodhouse1992, kirillov2001} extends classical mechanics into the quantum domain by quantizing symplectic manifolds; asymptotic quantization \cite{ashtekar1981, ashtekar1997, ashtekar2014}, developed by Ashtekar, focuses on quantizing the radiative modes of the gravitational field at null infinity; deformation quantization \cite{snyder1947, kontsevich2003} modifies the algebra of classical observables into a non-commutative structure to bridge classical and quantum mechanics; loop quantum gravity \cite{rovelli2008, ashtekar2021, rovelli2011b}, emphasizes a background-independent quantization of spacetime; and string theory \cite{becker2006, zwiebach2004}, which posits that fundamental particles are one-dimensional strings whose vibrations correspond to different particles, including gravitons. Despite their innovative approaches, each of these theories grapples with unique challenge: producing testable predictions.

A fundamental obstacle in developing a relativistically invariant quantum theory lies in the non-zero commutation relations between position and momentum operators, typically expressed as \cite{heisenberg1927}:
\begin{align}
[\hat{x}_i, \hat{p}_j] = \ii \hbar \delta_{ij}.
\end{align}
This relation is a cornerstone of the Heisenberg Uncertainty Principle (HUP), which asserts the impossibility of simultaneously determining a particle's position and momentum with arbitrary precision. Extensions and modifications of the HUP, such as the Generalized Uncertainty Principle (GUP) \cite{snyder1947, kempf1995, shah2023, maslowski2012} and those arising from the Moyal bracket \cite{moyal1949, kontsevich2003, sternheimer1998}, propose physics at very small scales, either by deforming the algebra of classical observables or faciliting a seamless transition between classical and quantum descriptions.

Building upon Dirac's pioneering work \cite{dirac1981, ashtekar1994}, physicists have endeavored to develop a quantization map that translates the Poisson bracket of classical mechanics into the quantum commutator. However, Groenewold's theorem \cite{groenewold1946, ali2005} highlights a significant impediment: there exists no quantization map that simultaneously maps all classical observables to quantum operators while preserving their commutation relations for all polynomials up to degree three. This theorem underscores an inherent incompatibility between classical and quantum mechanical descriptions at a fundamental level, complicating the pursuit of a unified theory.

Geometric quantization \cite{kostant1970, woodhouse1992, kirillov2001} offers a resolution by imposing conditions on the quantization of symplectic manifolds, ensuring that quantum operators derived from classical observables respect the underlying geometric and topological structures. This approach circumvents the direct issues associated with polynomial quantization. Additionally, the Wigner-Weyl scheme \cite{wigner1932, weyl1927} provides a systematic method for phase space quantization, utilizing polynomial ordering to resolve ambiguities in operator ordering. This method is pivotal for establishing a consistent quantization procedure for complex systems and aligns with the constraints imposed by Groenewold's theorem.

Despite these advancements, operator ordering schemes based on polynomial quantization \cite{dewitt1957} have faced criticism. In the conformal path integral formulation of quantum processes \cite{weiss1978}, often referred to as Feynman quantization, unjustified assumptions regarding ordering parameters persist, adhering to the McCoy formula \cite{mccoy1932}. For example, the operator $:\hat{x}\hat{p}^2:_\text{Weyl}$ is conventionally fixed at a constant value of $\frac{1}{3}$:
\begin{align}
:\hat{x}\hat{p}^2:_\text{Weyl} = \frac{1}{3} \hat{x}\hat{p}^2 + \frac{1}{3} \hat{p} \hat{x}\hat{p} + \frac{1}{3} \hat{p}^2 \hat{x},
\end{align}
which may not accurately capture the underlying physical reality.

In response to the quantization challenges, several semiclassical theories—hybrid models combining classical and quantum elements—have been proposed \cite{oppenheim2023, kafri2014, peres2004}. While some of these theories have been experimentally ruled out \cite{altamirano2018}, others, such as those based on the thermally-averaged stress-energy tensor \cite{becattini2015}, present a more viable framework. These models are grounded in relativistic hydrodynamics \cite{zubarev1979, van1982, frolov1987} and offer a means to describe quantum systems within curved spacetime environments.

Recent advances show that atomic clocks, functioning as precise quantum systems in relativistic fields, provide powerful tools for exploring quantum field effects in gravity. Gravitational time dilation, traditionally a classical concept, gains additional features in quantum mechanics, leading to observable effects like decoherence and phase shifts at the quantum level \cite{zych2011quantum, pikovski2015universal, pikovski2017time}. Experimental and theoretical studies \cite{paczos2024quantum, cepollaro2023gravitational, roura2021measuring, smith2020quantum, dkebski2024universality} reveal that gravitational quantum time dilation (GQTD) can alter coherence, interfere with clock states, and enhance quantum sensing. Delocalized superpositions in gravitational fields experience differential proper times, directly affecting clock observables. Our framework, while focusing on thermodynamic quantities like specific heat, naturally connects to this paradigm, linking radiative properties and temporal evolution of quantum states to quantum time dilation \cite{zych2017quantum}. This connection suggests that atomic clocks can test quantum effects in gravitational backgrounds, bridging relativistic quantum thermodynamics with quantum information in curved spacetime.

Currently, the foremost experimental approach for probing the quantum nature of gravity is the Bose-Marletto-Vedral (BMV) experiment \cite{bose2017, marletto2017}. This experiment involves two masses following superpositions of paths, with the goal of measuring gravity-induced entanglement. In its original formulation, the gravitational interaction between the masses is treated classically, governed by the quantized Newton potential $\hat{V} = - G \frac{m_1 m_2}{|\hat{x}_1 - \hat{x}_2|}$. Extensions of this setup to incorporate general relativity have been proposed \cite{martin2023}, and the necessity of integrating quantum degrees of freedom into relativistic information protocols has been emphasized \cite{perche2023}. Additionally, in a recent article, Tobar et al. proposed a gravito-phonoelectric analogue of the photoelectric effect, wherein ground-state quantum acoustic resonators, cooled to their fundamental mode, absorb individual gravitons and exhibit discrete energy changes akin to electron ejection \cite{tobar2024detecting}. Continuous monitoring of vibrational levels enables direct detection of quantum gravitational quanta, offering an experimental test of gravity’s quantum nature.

The approach presented in our work diverges from entanglement- or gravitational waves- based methods. Instead, we propose an experimental framework for testing the quantum effects in general relativity through the measurement of the heat capacity of a system. In Sec.~\ref{sec:ii} we present the setup, including the formulation of the relativistically invariant Hamiltonian and the canonical quantization procedure with arbitrary operator ordering. Sec.~\ref{sec:iii} provides a detailed derivation of the quantum Hamiltonian incorporating ordering parameters and operator commutation effects. In Sec.~\ref{sec:iv} we apply this framework to the thermodynamics of a gas in Rindler coordinates (focusing on temperatures above the Bose-Einstein condensation or Fermi thresholds) and we discuss perturbative calculations of specific heat, accounting for relativistic quantum corrections. Numerical analyses quantifying the dependence of specific heat on the ordering parameters are presented in Sec.~\ref{sec:v}. Finally, Sec.~\ref{sec:v} summarizes our conclusions and outlines prospects for experimental verification of the proposed framework.

Two key contributions emerge from our work. First, it proposes solution to the operator ordering problem within the relativistically invariant quantization framework. Second, it computes the specific heat for a bosonic system in Rindler coordinates, incorporating quantum corrections with potential experimental relevance. Our analysis incorporates both a semiclassical approximation for a Bose gas above the condensation temperature and a perturbative expansion utilizing Airy function solutions. This comprehensive theoretical framework allows for the derivation of corrections to the specific heat that transcend classical calculations, highlighting quantum dynamical effects that may be experimentally observable.

\section{Setup} \label{sec:ii}
Our study focuses on the quantum behavior of bosonic systems under relativistic and curved spacetime conditions, bridging various theoretical frameworks to address challenges in canonical quantization and experimental feasibility. The investigation spans both semiclassical and fully quantum regimes, employing an approach applicable above the critical condensation temperature.

A central objective is to scrutinize Dirac's proposal for relativistic canonical quantization \cite{dirac1981}, particularly concerning operator ordering ambiguities that emerge when quantizing systems in curved spacetimes \cite{suzuki1980, christodoulakis1986, anderson2012}. Building upon the insights of Khandelwal et al. \cite{khandelwal2020} and Dębski et al. \cite{debski2022}, this work formulates a relativistically invariant quantum Hamiltonian, rigorously resolves the ordering problem and enables the calculation of the specific heat of a quantum bosonic gas in Rindler coordinates.

The foundation of our study rests on the relativistically invariant \cite{einstein1915} proper mass of a particle with classical momentum $p_\mu$ in a gravitational background described by the metric $g^{\mu\nu}$:
\begin{align}
mc^2 = \sqrt{c^2 g^{\mu \nu} p_\mu p_\nu},
\end{align}
where $m$ represents the proper mass and $c$ the speed of light. Assuming that the mixed space-time components of the metric vanish and performing subsequent algebraic manipulations, the energy expression becomes:
\begin{align} \label{eq:energy_classical}
E = \sqrt{g_{00}} \sqrt{m^2 c^4 - c^2 g^{ij} p_i p_j}.
\end{align}
The primary objective is to quantize this classical energy expression. The quantization process entails the following steps:
\begin{enumerate}
    \item Transforming the Rest Energy: The rest energy of the particle is transformed into a rest quantum Hamiltonian $\hat{H}_0$. This Hamiltonian may describe internal degrees of freedom (such as a two-level system) or provide a more physically accurate representation of the mass term, e.g., a boson field in a cavity.
    \item Introducing Quantum Operators: Quantum position $\hat{x}_i$ and momentum $\hat{p}_i$ operators are introduced. In this model, the internal and external quantum degrees of freedom are decoupled, meaning $\hat{x}_i$ and $\hat{p}_i$ commute with $\hat{H}_0$.
    \item Assuming Higher-Order Corrections are Negligible: Higher-order corrections beyond second order in $\hat{p}$ and first order in $\hat{x}$ are assumed to be much smaller than the internal Hamiltonian. Subsequently, the metric tensor and square roots are expanded in a Taylor series to obtain the leading-order corrections to the Hamiltonian.
\end{enumerate}
In this framework, the back-reaction \cite{marto2021} of the quantum state on the metric tensor is neglected. The calculations are simplified by employing a diagonal metric tensor to describe a spherically symmetric system and considering only the radial position and momentum operators as non-zero. While generalizing this approach to more complex metrics is straightforward, such extensions are beyond the scope of this work. With the Hamiltonian established, the thermodynamics of the approximated relativistically invariant quantum operator are explored.

After deriving the relativistic quantum Hamiltonian, we proceed to investigate the specific heat of a system of particles governed by this operator. For simplicity and analytic tractability, we focus on Rindler spacetime and restrict our analysis to temperatures above the critical Bose and Fermi temperatures. This allows us to work within the Boltzmann statistical regime and avoid the complications of quantum statistical sums and condensation phenomena.

A semiclassical calculation—approximating the quantum partition function by an integral over phase space—yields a trivial result for the specific heat, independent of the system's structure. To capture meaningful corrections, we employ perturbation theory to compute the leading-order modification to the specific heat, which turns out to be proportional to the operator ordering parameter introduced in the Hamiltonian. We then numerically evaluate the magnitude of these corrections for physically realistic parameter values, demonstrating their potential observability.

\section{Relativistic Quantum Hamiltonian} \label{sec:iii}
For simplicity, let's assume that the particle's momentum and position operators align with the radial coordinate. Consequently, the only spacelike component of the metric tensor that multiplies a non-vanishing term in Eq.\eqref{eq:energy_classical} is $g^{rr}$. Hereafter, subscripts $r$ will be omitted for the position and momentum operators.

    First, we notice that, when we make the metric tensor a function of operator of position, there appears an amibiguity in ordering of multiplication of the square roots in the Eq.\eqref{eq:energy_classical}. By introducing the time-dilation operator $\sqrt{g_{00}(r+\hat{x})}$, where $r$ denotes the reference position and $\hat x$ represents the quantum displacement from that point, we can express the relativistically invariant Hermitian Hamiltonian of a particle with quantum degrees of freedom as follows:
        \begin{align} \label{eq:h-inv-1}
            \hat{H} = \gamma \sqrt{g_{00}(r+\hat{x})} \hat{H}_p + \gamma^*\hat H_p\sqrt{g_{00}(r+\hat{x})},
        \end{align}
    with $\gamma$ being a complex parameter such that $2 \text{Re}\gamma = 1$ in order to satisfy the classical limit of commuting position and momentum and $\hat{H}_p$ is the kinetic Hamiltonian derived from Eq.\eqref{eq:energy_classical}:
    \begin{align} \label{eq:h-inv-p}
        \hat{H}_p = \sqrt{ \hat{H_0}^{2} - c^2 : g^{rr}(r+\hat{x}) \hat{p}^2:},        
    \end{align} 
    where $:\cdot:$ is the general ordering defined below (see Eqs.\eqref{eq:or1}-\eqref{eq:xp2-2}, that arises from the commutation relation $[\hat{p},\hat{x}]=-\ii\hbar$.

    As the metric component now depends on the position operator, but the metric tensor is defined only for scalar variables, a covariant Taylor series expansion is necessary with $\frac{\hat{x}\phi}{r\phi} \ll 1$:
        \begin{align} \label{eq:or1}
            : g^{rr}(r+\hat{x}) \hat{p}^2 : \; = g^{rr}(r) \hat{p}^2 + \sum_{k=1}^\infty \partial_r^k g^{rr}(r) : \hat{x}^k \hat{p}^2 :,
        \end{align}
    where $: \hat{x}^k \hat{p}^2 :$ equals:
        \begin{align} \label{eq:xp2-2}
            : \hat{x}^k \hat{p}^2 : \;  = \hat{x}^k \hat{p}^2 -\ii (1-2\ii\operatorname{Im} \alpha) k \hbar \hat{x}^{k-1} \hat{p} - \alpha k(k-1)\hbar^2 \hat{x}^{k-2},
        \end{align}
    that is derived in detail in Appendix \ref{A}. What is important at this point is that usually (in the Weyl ordering), it is assumed that $\alpha = \frac{1}{3}$, and straighforwardly, we can see that in principle, this parameter may have another value, even an imaginary one.

    Substituting all, expanding in the Taylor serieses, ommiting higher order terms, as stated in the Setup, and detailed in Appendix \ref{A}, we arrive at:
        \begin{align} \label{eq:h-inv-final}
            \hat{H} =  \hat{H_0} \sqrt{g_{00}} \Bigg[&1+\frac{g'_{00}}{2g_{00}} \hat{x} \nonumber 
            \\&- \frac{ c^2} {2\hat{H_0}^2} \left( g^{rr} + \left(g'^{rr} + \frac{ g'_{00}g^{rr}}{2g_{00}} \right) \hat{x} \right) \hat{p}^2 \nonumber \\
            &+\frac{ \ii \hbar c^2} {2\hat{H_0}^2} \left((1-2\ii\operatorname{Im} \alpha) g'^{rr} +\gamma^*\frac{ g'_{00}g^{rr}}{g_{00}} \right) \hat{p}  \Bigg],
        \end{align}
    where $g^{\mu\nu}(r) \equiv g^{\mu\nu}$ and $\partial_r g^{\mu\nu}(r) \equiv g'^{\mu\nu}$.
    Eq.\eqref{eq:h-inv-final} represents a first-order correction to the relativistic Hamiltonian of a system possessing quantum internal and radial dynamic degrees of freedom.

    Equation~\eqref{eq:h-inv-final} constitutes a central result of this work, encapsulating the first-order relativistic quantum correction to the Hamiltonian of a particle in a curved spacetime with both internal and external degrees of freedom. Notably, this expression introduces two ordering parameters: $\alpha$, governing the ordering of operators associated with the spatial part of the metric, and $\gamma$, controlling the ordering linked to the temporal metric component. This distinction enables a systematic exploration of the contributions from different operator orderings to physical observables. For metrics that are nonlinear functions of position, our formalism allows for a Taylor expansion of the metric operator up to any desired order. Consequently, successive corrections to the Hamiltonian can be computed, arising from the nontrivial commutation relations between momentum operators and metric components expressed as functions of position operators. This structure provides a foundation for perturbative calculations of quantum thermodynamic properties, sensitive to the operator ordering ambiguity inherent in canonical quantization on curved backgrounds.

\section{Thermodynamics of a Quantum Boltzmann Gas in Rindler Spacetime}  \label{sec:iv}
In this section, we analyze Eq.~\eqref{eq:h-inv-final} for the specific case of Rindler spacetime, which models a uniformly accelerating frame and locally mimics a gravitational field near a black hole horizon. In this setting, the square root of the temporal metric component varies linearly with distance from the reference hyperbola, so the only deviation from Minkowski spacetime is the local time dilation: $\sqrt{g_{00}(\hat{x})} = 1 + \frac{g \hat{x}}{c^2}$, without reference to any central point, ie. $r = 0$.

Starting from Eq.~\eqref{eq:h-inv-1}, and without assuming the small-energy approximation $\frac{\hat{p}^2}{mc^2} \ll 1$, the relativistic quantum Hamiltonian in the Rindler frame takes the form
\begin{align} \label{eq:ham-ful-rindler}
    \hat{H} =\hat{H}_0 \Bigg[\gamma^* \left(1+\frac{g\hat{x}}{c^2}\right)\sqrt{1+\frac{c^2\hat{p}^2}{H_0^2}} + \gamma \sqrt{1+\frac{c^2\hat{p}^2}{H_0^2}} \left(1+\frac{g\hat{x}}{c^2}\right) \Bigg].
\end{align}

Our goal is to commute the square-root operator involving momentum to the right-hand side of the position operator. However, this operation is not straightforward. To proceed, consider an arbitrary function of the momentum operator, $f(\hat{p})$, and compute its commutator with the position operator. In the momentum representation, we have the well-known identity:
\begin{align} \label{eq:comu-f}
    [f(\hat{p}), \hat{x}] &= - \ii \hbar \pdv{f(\hat{p})}{\hat{p}} .
\end{align}

Using Eq.~\eqref{eq:comu-f}, the Hamiltonian in Eq.~\eqref{eq:ham-ful-rindler} can be rewritten as
\begin{align} \label{eq:h-rind-prec}
    \hat{H} =\hat{H}_0 \Bigg[&\left(1+\frac{g\hat{x}}{c^2}\right)\sqrt{1+\frac{c^2\hat{p}^2}{\hat{H}_0^2}} -\ii \gamma \frac{\hbar g}{\hat{H}_0^2} \frac{\hat{p}}{\sqrt{1+\frac{c^2\hat{p}^2}{\hat{H}_0^2}}}\Bigg],
\end{align}
which constitutes a non-approximated expression for the Hamiltonian of a relativistic quantum system with dynamical degrees of freedom in Rindler coordinates.

Setting $H_0=mc^2$ and taking the low-energy limit of the full relativistic Hamiltonian, we obtain the simplified expression:
\begin{align} \label{eq:h-rind-low}
    \hat{H} =mc^2\left(1+\frac{g\hat{x}}{c^2}\right)\left(1+\frac{\hat{p}^2}{2m^2c^2}\right) - \ii \gamma \frac{ \hbar g \hat{p}}{mc^2}.
\end{align}
This form describes a particle in a linear gravitational potential with relativistic corrections included perturbatively.

\begin{figure}[hbtp]
\centering
\includegraphics[width=0.47\textwidth]{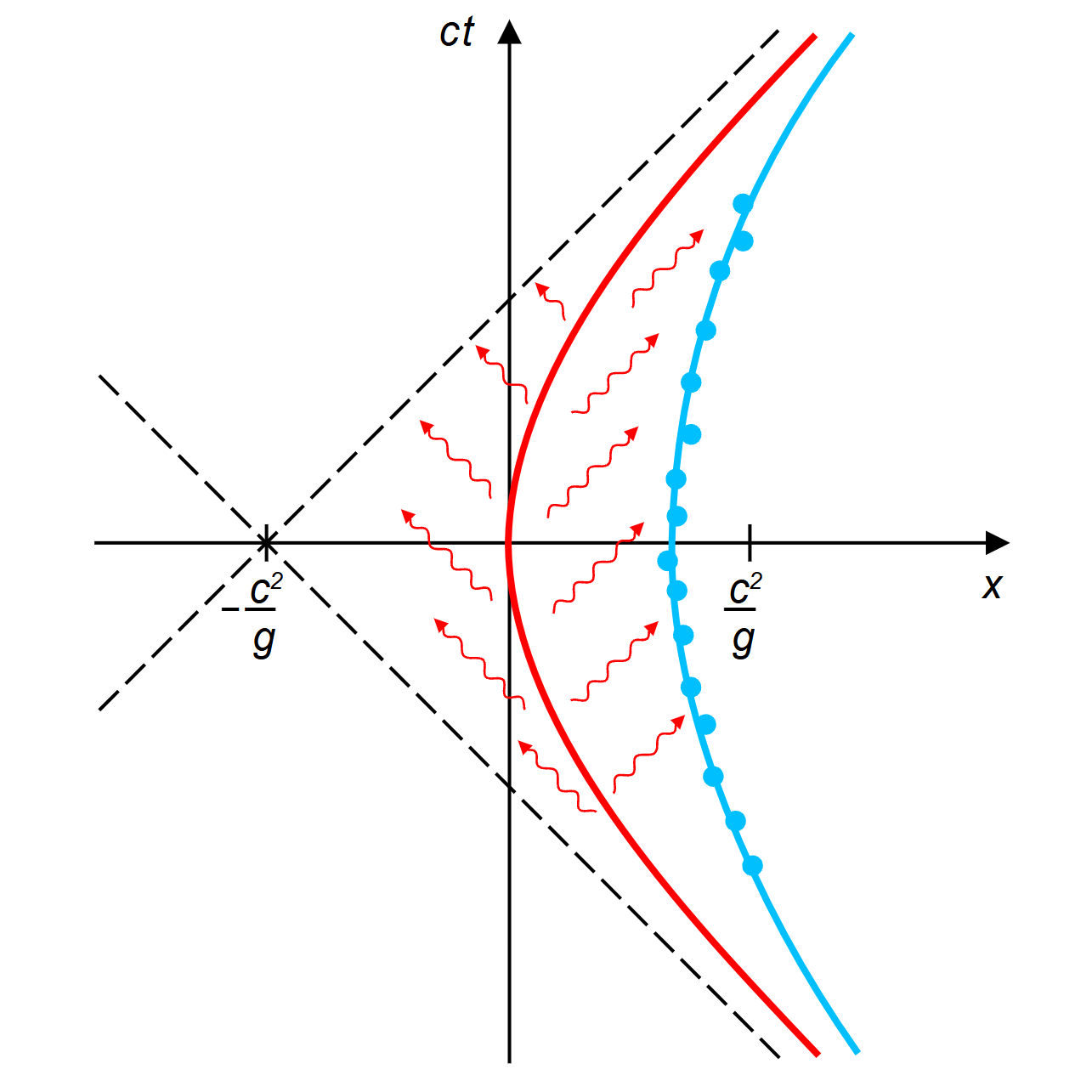}
\caption{\justifying \fontsize{9pt}{11pt}\selectfont The right Rindler wedge is depicted, with the red curve marking the reference hyperbola of constant acceleration $g$. The event horizon, at a proper distance $x=-\frac{c^2}{g}$ from the hyperbola, is shown as the dashed black line light cone. Radiation emitted by the heat bath on the hyperbola maintains thermal equilibrium throughout the wedge; however, gravitational redshift causes the local temperature to vary according to the Tolman–Ehrenfest relation. The blue line indicates our system setup at the proper distance $x$ from the reference hyperbola, and the dots represent gas particles.}
\label{fig:hb}
\end{figure}

In the following thermodynamic analysis, $T$ refers to the local proper temperature, so when we study $C_V$ as a function of the distance from the heat bath, we must enforce relativistic thermal equilibrium via the classical Tolman–Ehrenfest relation \cite{tolman1928, tolman1930}:
\begin{align}
T(x)\sqrt{g_{00}(x)} = \Theta_\text{hb},
\end{align}
which prescribes how the local temperature $T(x)$ varies in a stationary spacetime in equilibrium with a bath at temperature $\Theta_\text{hb}$. In Rindler coordinates, $\sqrt{g_{00}(0)} = 1$, so by placing the heat bath on the reference hyperbola of acceleration $g$, we can examine the specific heat profile as a function of the classical distance $x > -c^2/g$. This setup is shown schematically in Fig.~\ref{fig:hb}.

Before proceeding, let us comment on the physical validity of the approximations employed. The assumption of the low-energy limit is justified provided that $T \ll \frac{mc^2}{k_\text{B}}$, a condition that is safely satisfied within the temperature range considered in this work. The simplest method for calculating the specific heat is the semiclassical approximation, which amounts to assuming that position and momentum operators effectively commute and replacing the sum over discrete energy levels by an integral over phase space. However, to ensure the validity of this approximation, the temperature must satisfy $T \gg \Delta E$, where $\Delta E$ denotes the characteristic energy scale of the system. As shown in Appendix~\ref{B}, in this limit the semiclassical partition function yields a trivial result for the specific heat:
\begin{align}
C_V(T, N) = k_\text{B} N,
\end{align}
which provides no meaningful insight into quantum or relativistic corrections.

In our regime of interest, however, the condition for the validity of the semiclassical approximation:
\begin{align}
T_\text{semiclassical} \gg \frac{\Delta E}{k_\text{B}} \propto \frac{\left( \hbar^2 m g^2 \right)^{1/3}}{k_\text{B}},
\end{align}
is not satisfied. This emphasizes the necessity of adopting a fully quantum treatment, as pursued in the subsequent analysis.

Consequently, we interpret the Hamiltonian of Eq.~\eqref{eq:h-rind-low} as a standard nonrelativistic Hamiltonian with a linear gravitational potential, perturbed by relativistic terms. The leading-order energy levels are governed by the unperturbed part, whose eigenfunctions are expressed in terms of Airy functions. Building on this, we present a detailed perturbative analysis: we compute the eigenenergies up to second order and the eigenfunctions up to first order using time-independent perturbation theory. This approach allows us to extract the quantum corrections to specific heat beyond the semiclassical approximation.

As shown in the Appendix \ref{C}, the eigenenergies of unperturbed Hamiltionian $\hat{H}_0 = mc^2 + \frac{\hat{p}^2}{2m} + mg \hat{x}$ are expressed in terms of Airy functions \cite{vallee2010airy, bieniek1977uniform}:
\begin{align}
    E_n^{(0)} &= mc^2 + \epsilon_n,
\end{align}
where the energy shift equals $\epsilon_n = a_n mgL$ with $a_n$ being one of the zeros of the Airy function $\Ai{-a_n} = 0$ and the characteristic length $L = \left( \frac{\hbar^2}{2m^2g} \right)^{1/3}$.

The perturbative derivation of the energy spectrum with the perturbation Hamiltonian $\hat{H}'=\frac{g\hat{x}\hat{p}^2}{2mc^2} - \ii \gamma \frac{\hbar g \hat{p}}{mc^2}$ yield in the first order perturbation:
\begin{align} 
    E_n^{(1)} = \frac{mg^2L^2}{c^2} \left(a_n \ppbraket{n}{\hat{\xi}}{n} - \ppbraket{n}{\hat{\xi}^2}{n} \right),
\end{align}
where $\hat\xi = \frac{\hat{x}}{L}$, while the second-order correction equals: 
\begin{align} 
    E_n^{(2)} = \frac{mg^3L^3}{c^4} \sum_{k \neq n} \frac{\left| \left(a_n(1-\gamma) + \gamma a_k\right) \ppbraket{k}{\hat{\xi}}{n} - \ppbraket{k}{\hat{\xi}^2}{n}\right|^2}{a_n-a_k}.
\end{align}
Assuming an orthonormal basis, the matrix elements $\ppbraket{k}{\hat{\xi}}{n}$ and $\ppbraket{k}{\hat{\xi}^2}{n}$ can be approximated by derivatives of Kronecker deltas, as shown in Appendix~\ref{C}.

From the full energy spectrum $E_n = E_n^{(0)}+E_n^{(1)}+E_n^{(2)}$, the canonical partition function at temperature $T$ is $Z = \sum_n \ee^{-\beta E_n}$, where $\beta^{-1} = k_\text{B} T$. Utilizing the Boltzmann statistics, the specific heat per particle at constant volume is then computed from \cite{tong_statistical_physics}: 
\begin{align} \label{eq:spec-he}
    \frac{C_V}{N} = \frac{\partial}{\partial T} \left( \frac{\tbraket{E}}{Z} \right) = \frac{1}{k_B T^2} \left( \tbraket{E^2} - \tbraket{E}^2 \right), 
\end{align} 
where the thermal averages $\tbraket{E} = \sum_n E_n \ee^{-\beta E_n}$ and $\tbraket{E^2} = \sum_n E_n^2 \ee^{-\beta E_n}$ are taken with respect to the full energy spectrum.

Our analysis proceeds by substituting the explicit expressions for the zeroth- and first-order corrections to the energy spectrum (i.e., $E_n^{(0)}$ and $E_n^{(1)}$) into the canonical partition function, leading to the reference specific heat $C_V^0$. These terms do not depend on the ordering parameter $\gamma$ and can therefore be interpreted as arising from a semiclassical approximation in which position and momentum operators effectively commute. In contrast, the specific heat $C_V$, calculated up to second-order corrections to the energy spectrum (i.e., including $E_n^{(0)}$, $E_n^{(1)}$, and $E_n^{(2)}$), encapsulates the contribution of ordering effects via $\gamma = \frac{1}{2} + \ii\operatorname{Im} \gamma$. By varying the imaginary part of the ordering parameter $\gamma$, we can systematically examine the sensitivity of specific heat to different operator orderings. Successive perturbative corrections are developed using the orthogonality relations of the Airy function eigenbasis, as detailed in Appendix~\ref{C}, where matrix elements are expressed in terms of Kronecker delta derivatives. This formalism allows us to isolate the influence of operator ordering parameters on the thermodynamic behavior. Our numerical analyses further quantify these dependencies, enabling precise predictions of specific heat as a function of local temperature, under the assumption of thermal equilibrium with a local heat bath at each spatial point.

This perturbative framework allows for the systematic inclusion of relativistic corrections to the thermodynamic behavior of quantum systems in a linear gravitational potential and will serve as the basis for numerical evaluation of specific heat of a quantum gas under strong acceleration. Observing these corrections experimentally could provide insights into the interplay between quantum mechanics and general relativity, offering a approach to determination of ordering parameters in relativistic canonical quantization.

\section{Numerical Analysis of Specific Heat}  \label{sec:v}
We analyze Eq.\eqref{eq:spec-he} in the parameter regime where quantum-relativistic corrections to specific heat are clearly visible, following the procedure outlined in Sec.\ref{sec:iv}. We plot the specific heat per particle as a function of the local temperature and compare three spectra: (i) the zeroth- plus first-order result that neglects the $E_n^{(2)}$ term, (ii) the fully corrected Weyl-ordered spectrum, and (iii) the spectrum that includes an additional imaginary contribution to the ordering parameter $\gamma=\tfrac12+\ii\operatorname{Im}\gamma$.

\begin{figure}[hbtp]
\centering
\includegraphics[width=0.47\textwidth]{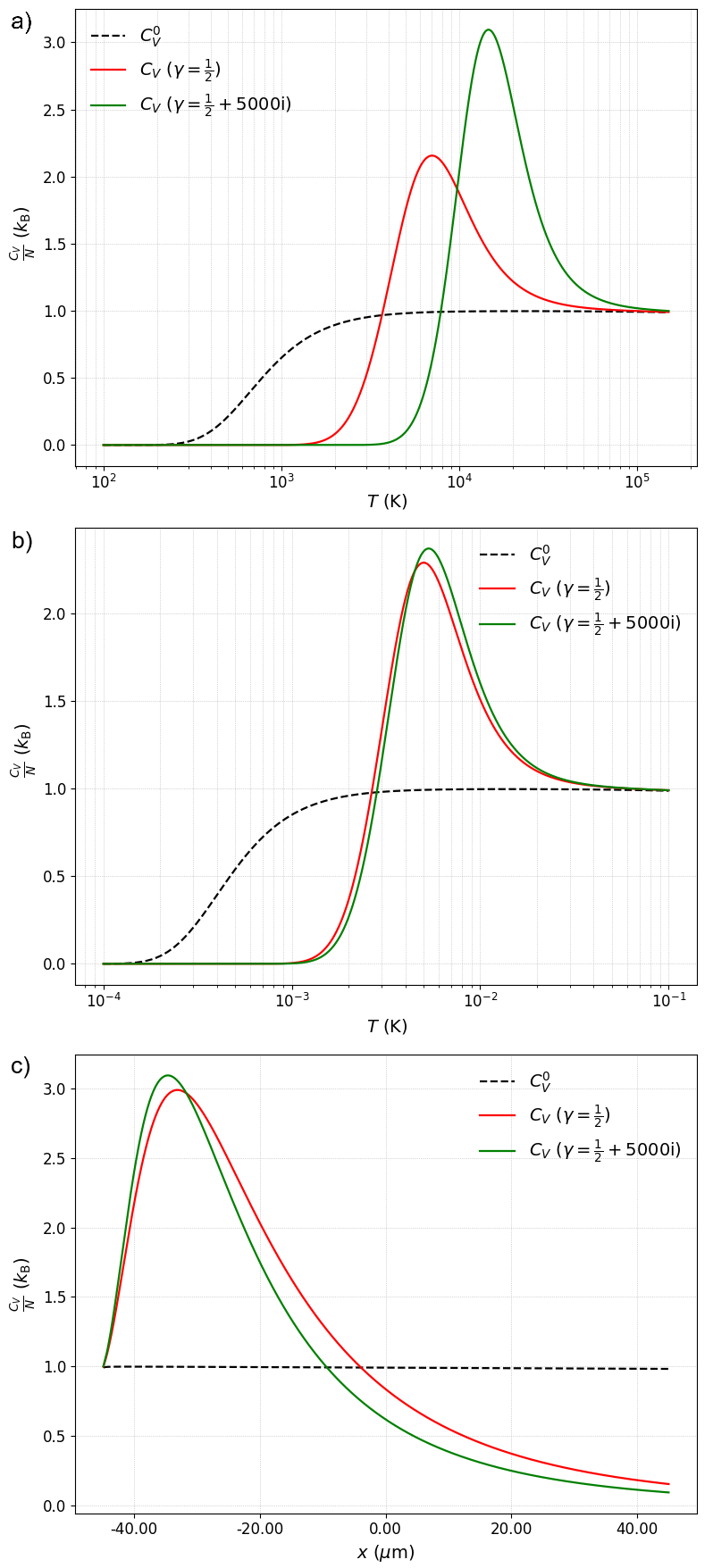}
\caption{\justifying \fontsize{9pt}{11pt}\selectfont Specific heat per particle for three scenarios, panels (a)–(c): black dashed—spectrum to first order only; red—full Weyl-ordered second-order correction; green—additional imaginary part of $\gamma$. (a) Electrons in a uniformly accelerated frame: second-order and $\operatorname{Im}\gamma$ terms markedly reshape the $C_V(T)$ curve. (b) Ultra-light neutrinos in a strong static gravitational field: the same corrections are visible but smaller, reflecting the lower mass scale. (c) The electron system is revisited, now plotting $C_V(x)$ against the distance from the reference hyperbola—where the heat bath is placed—so that the Tolman–Ehrenfest relation renders the local temperature position-dependent and ordering effects blend with relativistic red-shift.}
\label{fig:cv}
\end{figure}

Figure~\ref{fig:cv}a) concerns electrons of mass $m_\text{e}=9.1\times10^{-31}\text{kg}$ subjected to a uniform electric-field acceleration $a=2\times10^{21} \frac{\text{m}}{\text{s}^{2}}$, a value that can be approached in state-of-the-art laser accelerators.  The ordering-dependent correction is most pronounced at intermediate temperatures and its magnitude increases with stronger acceleration.

Figure~\ref{fig:cv}b) addresses ultra-light particles, taking $m_\mu=8\times10^{-37}\text{kg}$ as a representative neutrino mass and $a=10^{15}\frac{\text{m}}{\text{s}^{2}}$ as the proper acceleration expected near the Schwarzschild horizon of an extreme black hole. Although the qualitative temperature dependence mirrors the electron case, the specific-heat scale is greatly reduced because it is proportional to the particle mass and the proper acceleration.

Finally, Figure~\ref{fig:cv}c) illustrates the influence of the Tolman–Ehrenfest relation on the specific-heat profile. We fix the bath temperature at $\Theta_\text{hb}=7\times10^{3}\text{K}$ on the reference hyperbola and track the specific heat of electrons as a function of the classical distance $x$ measured along the Rindler direction. Because the local temperature scales as $T(x)=\sqrt{g_{00}(x)}^{-1}\Theta_\text{hb}$, second-order corrections become increasingly significant with distance, shifting both the location and the height of the specific-heat maxima. This spatial dependence may provide an experimental handle for inferring ordering parameters in precision calorimetric setups.

The characteristic peak–dip–plateau structure of the specific‐heat curve we obtain closely resembles the behavior reported in Ref.~\cite{rouabhia2023effect}, where the thermodynamic properties of a one‐dimensional Klein–Gordon oscillator in Rindler spacetime were derived via a zeta‐function–regularized partition function obtained through the Cahen–Mellin transform. In their setup, spin‐zero bosons subject to a linear acceleration $g$ experience modified energy levels from the Rindler‐space Klein–Gordon potential, and all thermodynamic quantities—including the specific heat—were analyzed as functions of $g$, whereas our analysis further explores the additional dependence on operator‐ordering parameters and Tolman-Ehrenfest relation.

Our three examples indicate a plausible experimental route to determine the ordering parameter $\gamma$ from precision measurements of specific heat. Extending the same methodology to curved spacetimes, such as the Schwarzschild metric, would enable tests of operator-ordering ambiguities in genuinely relativistic quantum systems. In particular, the spatial-ordering parameter $\alpha$ introduced earlier should, in principle, be accessible through analogous calorimetric studies.

\section{Conclusions} \label{sec:vi}
A central question emerges: Are the theoretical predictions presented here experimentally accessible? In the case of electrons undergoing uniform acceleration, the predicted corrections to the specific heat due to the operator ordering parameter appear feasible within the considered acceleration regimes. However, detecting analogous effects in gravitational settings remains a far greater challenge. For neutrinos, while modifications to their specific heat might occur near extreme astrophysical objects such as dense black holes, their purely gravitational interactions make laboratory manipulation virtually impossible.

Extending this framework to Bose and Fermi statistics would enable the investigation of relativistic corrections to thermodynamic quantities at ultra-low temperatures, using Hydrogen-1 or Sodium-23 atoms under more experimentally accessible acceleration regimes. Modern experimental techniques might already allow for the realization of such conditions. Temperatures as low as $T = 10^{-12}\,\text{K}$ have been achieved~\cite{deppner2021collective} and accelerator rings have been used to accelerate Bose–Einstein condensates (BECs)~\cite{pandey2019hypersonic}. 

However, to support such ambitious experimental efforts, a robust theoretical framework is essential. The interplay between mechanical stress, thermal regulation under extreme acceleration, and the fragile nature of trapped BECs calls for highly precise experimental design. Mitigating the destabilizing effects of acceleration requires careful control, and additional phenomena such as Landau–Zener tunneling~\cite{wu2003superfluidity, choi1999bose}, dynamical instabilities~\cite{burger2001superfluid}, and energy shifts due to acceleration must also be considered. Accurate measurements of specific heat—potentially through quantum calorimetry techniques~\cite{newsome2004relaxation}—are a critical component of any such verification.

A notable step toward experimental realization is the work by P\"otting et al.~\cite{potting2001coherent}, who propose a method for accelerating BECs using a frequency-chirped optical lattice formed by counterpropagating laser beams. Their theoretical model describes the acceleration process and analyzes the condensate’s stability. Numerical simulations confirm that BECs can remain stable under accelerations up to $a = 10^5\, \text{m/s}^2$, offering valuable insight into the behavior of ultracold quantum gases in accelerated frames. This pioneering effort paves the way for future experiments that probe quantum systems under non-inertial conditions.

The main result of this work is the formulation of a relativistic quantum Hamiltonian that incorporates both dynamical degrees of freedom and an internal system Hamiltonian, respecting covariant structure. Using this Hamiltonian, we derived the specific heat of a bosonic quantum gas in Rindler coordinates and identified measurable corrections arising from nontrivial commutators between position and momentum. Although subtle, these quantum corrections lie within the sensitivity of state-of-the-art experiments~\cite{collings1986low}, suggesting their potential observability.

If confirmed experimentally, the dependence of the specific heat of a Boltzmann gas on gravitational acceleration would offer direct evidence for operator-ordering effects—potentially deviating from Weyl ordering—and provide insights into the quantization procedure in curved spacetime. This analysis is situated within a first-quantization framework, without invoking quantum field theory or gravitons. Nonetheless, it offers a route for probing quantum phenomena in gravitational backgrounds, complementing other approaches such as entanglement-based tests in general relativistic settings \cite{bose2017, marletto2017, martin2023}.

In summary, this study advances the theoretical understanding of quantum systems in relativistic and curved spacetimes. By addressing the operator ordering ambiguity in relativistic quantization and calculating the corresponding thermodynamic corrections to specific heat, we provide a framework that may assist future experimental studies of quantum effects in gravitational backgrounds.

\section*{ACKNOWLEDGEMENT}
The authors would like to thank Prof. Andrzej Dragan for insightful discussions and valuable suggestions throughout the development of this work.

The data that support the findings of this article are openly available \cite{code}.

\bibliography{bibliography}

\begin{thebibliography}{76}%
\makeatletter
\providecommand \@ifxundefined [1]{%
 \@ifx{#1\undefined}
}%
\providecommand \@ifnum [1]{%
 \ifnum #1\expandafter \@firstoftwo
 \else \expandafter \@secondoftwo
 \fi
}%
\providecommand \@ifx [1]{%
 \ifx #1\expandafter \@firstoftwo
 \else \expandafter \@secondoftwo
 \fi
}%
\providecommand \natexlab [1]{#1}%
\providecommand \enquote  [1]{``#1''}%
\providecommand \bibnamefont  [1]{#1}%
\providecommand \bibfnamefont [1]{#1}%
\providecommand \citenamefont [1]{#1}%
\providecommand \href@noop [0]{\@secondoftwo}%
\providecommand \href [0]{\begingroup \@sanitize@url \@href}%
\providecommand \@href[1]{\@@startlink{#1}\@@href}%
\providecommand \@@href[1]{\endgroup#1\@@endlink}%
\providecommand \@sanitize@url [0]{\catcode `\\12\catcode `\$12\catcode `\&12\catcode `\#12\catcode `\^12\catcode `\_12\catcode `\%12\relax}%
\providecommand \@@startlink[1]{}%
\providecommand \@@endlink[0]{}%
\providecommand \url  [0]{\begingroup\@sanitize@url \@url }%
\providecommand \@url [1]{\endgroup\@href {#1}{\urlprefix }}%
\providecommand \urlprefix  [0]{URL }%
\providecommand \Eprint [0]{\href }%
\providecommand \doibase [0]{https://doi.org/}%
\providecommand \selectlanguage [0]{\@gobble}%
\providecommand \bibinfo  [0]{\@secondoftwo}%
\providecommand \bibfield  [0]{\@secondoftwo}%
\providecommand \translation [1]{[#1]}%
\providecommand \BibitemOpen [0]{}%
\providecommand \bibitemStop [0]{}%
\providecommand \bibitemNoStop [0]{.\EOS\space}%
\providecommand \EOS [0]{\spacefactor3000\relax}%
\providecommand \BibitemShut  [1]{\csname bibitem#1\endcsname}%
\let\auto@bib@innerbib\@empty
\bibitem [{\citenamefont {Peskin}(2018)}]{peskin2018}%
  \BibitemOpen
  \bibfield  {author} {\bibinfo {author} {\bibfnamefont {M.~E.}\ \bibnamefont {Peskin}},\ }\href@noop {} {\emph {\bibinfo {title} {An introduction to quantum field theory}}}\ (\bibinfo  {publisher} {CRC press},\ \bibinfo {year} {2018})\BibitemShut {NoStop}%
\bibitem [{\citenamefont {Feynman}(2006)}]{feynman2006}%
  \BibitemOpen
  \bibfield  {author} {\bibinfo {author} {\bibfnamefont {R.~P.}\ \bibnamefont {Feynman}},\ }\href@noop {} {\emph {\bibinfo {title} {QED: The strange theory of light and matter}}}\ (\bibinfo  {publisher} {Princeton University Press},\ \bibinfo {year} {2006})\BibitemShut {NoStop}%
\bibitem [{\citenamefont {Feynman}(2018)}]{feynman2018}%
  \BibitemOpen
  \bibfield  {author} {\bibinfo {author} {\bibfnamefont {R.}~\bibnamefont {Feynman}},\ }\href@noop {} {\emph {\bibinfo {title} {Feynman lectures on gravitation}}}\ (\bibinfo  {publisher} {CRC press},\ \bibinfo {year} {2018})\BibitemShut {NoStop}%
\bibitem [{\citenamefont {Zee}(2010)}]{zee2010}%
  \BibitemOpen
  \bibfield  {author} {\bibinfo {author} {\bibfnamefont {A.}~\bibnamefont {Zee}},\ }\href@noop {} {\emph {\bibinfo {title} {Quantum field theory in a nutshell}}},\ Vol.~\bibinfo {volume} {7}\ (\bibinfo  {publisher} {Princeton university press},\ \bibinfo {year} {2010})\BibitemShut {NoStop}%
\bibitem [{\citenamefont {Kostant}(1970)}]{kostant1970}%
  \BibitemOpen
  \bibfield  {author} {\bibinfo {author} {\bibfnamefont {B.}~\bibnamefont {Kostant}},\ }\href@noop {} {\bibfield  {journal} {\bibinfo  {journal} {Lect. Notes in Math.}\ }\textbf {\bibinfo {volume} {170}},\ \bibinfo {pages} {87} (\bibinfo {year} {1970})}\BibitemShut {NoStop}%
\bibitem [{\citenamefont {Woodhouse}(1992)}]{woodhouse1992}%
  \BibitemOpen
  \bibfield  {author} {\bibinfo {author} {\bibfnamefont {N.~M.~J.}\ \bibnamefont {Woodhouse}},\ }\href@noop {} {\emph {\bibinfo {title} {Geometric quantization}}}\ (\bibinfo  {publisher} {Oxford university press},\ \bibinfo {year} {1992})\BibitemShut {NoStop}%
\bibitem [{\citenamefont {Kirillov}(2001)}]{kirillov2001}%
  \BibitemOpen
  \bibfield  {author} {\bibinfo {author} {\bibfnamefont {A.~A.}\ \bibnamefont {Kirillov}},\ }in\ \href@noop {} {\emph {\bibinfo {booktitle} {Dynamical Systems IV: Symplectic Geometry and its Applications}}}\ (\bibinfo  {publisher} {Springer},\ \bibinfo {year} {2001})\ pp.\ \bibinfo {pages} {139--176}\BibitemShut {NoStop}%
\bibitem [{\citenamefont {Ashtekar}(1981)}]{ashtekar1981}%
  \BibitemOpen
  \bibfield  {author} {\bibinfo {author} {\bibfnamefont {A.}~\bibnamefont {Ashtekar}},\ }\href@noop {} {\bibfield  {journal} {\bibinfo  {journal} {Physical Review Letters}\ }\textbf {\bibinfo {volume} {46}},\ \bibinfo {pages} {573} (\bibinfo {year} {1981})}\BibitemShut {NoStop}%
\bibitem [{\citenamefont {Ashtekar}\ \emph {et~al.}(1997)\citenamefont {Ashtekar}, \citenamefont {Bi{\v{c}}{\'a}k},\ and\ \citenamefont {Schmidt}}]{ashtekar1997}%
  \BibitemOpen
  \bibfield  {author} {\bibinfo {author} {\bibfnamefont {A.}~\bibnamefont {Ashtekar}}, \bibinfo {author} {\bibfnamefont {J.}~\bibnamefont {Bi{\v{c}}{\'a}k}},\ and\ \bibinfo {author} {\bibfnamefont {B.~G.}\ \bibnamefont {Schmidt}},\ }\href@noop {} {\bibfield  {journal} {\bibinfo  {journal} {Physical Review D}\ }\textbf {\bibinfo {volume} {55}},\ \bibinfo {pages} {669} (\bibinfo {year} {1997})}\BibitemShut {NoStop}%
\bibitem [{\citenamefont {Ashtekar}(2014)}]{ashtekar2014}%
  \BibitemOpen
  \bibfield  {author} {\bibinfo {author} {\bibfnamefont {A.}~\bibnamefont {Ashtekar}},\ }\href@noop {} {\bibfield  {journal} {\bibinfo  {journal} {arXiv preprint arXiv:1409.1800}\ } (\bibinfo {year} {2014})}\BibitemShut {NoStop}%
\bibitem [{\citenamefont {Snyder}(1947)}]{snyder1947}%
  \BibitemOpen
  \bibfield  {author} {\bibinfo {author} {\bibfnamefont {H.~S.}\ \bibnamefont {Snyder}},\ }\href@noop {} {\bibfield  {journal} {\bibinfo  {journal} {Physical Review}\ }\textbf {\bibinfo {volume} {71}},\ \bibinfo {pages} {38} (\bibinfo {year} {1947})}\BibitemShut {NoStop}%
\bibitem [{\citenamefont {Kontsevich}(2003)}]{kontsevich2003}%
  \BibitemOpen
  \bibfield  {author} {\bibinfo {author} {\bibfnamefont {M.}~\bibnamefont {Kontsevich}},\ }\href@noop {} {\bibfield  {journal} {\bibinfo  {journal} {Letters in Mathematical Physics}\ }\textbf {\bibinfo {volume} {66}},\ \bibinfo {pages} {157} (\bibinfo {year} {2003})}\BibitemShut {NoStop}%
\bibitem [{\citenamefont {Rovelli}(2008)}]{rovelli2008}%
  \BibitemOpen
  \bibfield  {author} {\bibinfo {author} {\bibfnamefont {C.}~\bibnamefont {Rovelli}},\ }\href@noop {} {\bibfield  {journal} {\bibinfo  {journal} {Living reviews in relativity}\ }\textbf {\bibinfo {volume} {11}},\ \bibinfo {pages} {1} (\bibinfo {year} {2008})}\BibitemShut {NoStop}%
\bibitem [{\citenamefont {Ashtekar}\ and\ \citenamefont {Bianchi}(2021)}]{ashtekar2021}%
  \BibitemOpen
  \bibfield  {author} {\bibinfo {author} {\bibfnamefont {A.}~\bibnamefont {Ashtekar}}\ and\ \bibinfo {author} {\bibfnamefont {E.}~\bibnamefont {Bianchi}},\ }\href@noop {} {\bibfield  {journal} {\bibinfo  {journal} {Reports on Progress in Physics}\ }\textbf {\bibinfo {volume} {84}},\ \bibinfo {pages} {042001} (\bibinfo {year} {2021})}\BibitemShut {NoStop}%
\bibitem [{\citenamefont {Rovelli}(2011)}]{rovelli2011b}%
  \BibitemOpen
  \bibfield  {author} {\bibinfo {author} {\bibfnamefont {C.}~\bibnamefont {Rovelli}},\ }\href@noop {} {\bibfield  {journal} {\bibinfo  {journal} {Classical and Quantum Gravity}\ }\textbf {\bibinfo {volume} {28}},\ \bibinfo {pages} {114005} (\bibinfo {year} {2011})}\BibitemShut {NoStop}%
\bibitem [{\citenamefont {Becker}\ \emph {et~al.}(2006)\citenamefont {Becker}, \citenamefont {Becker},\ and\ \citenamefont {Schwarz}}]{becker2006}%
  \BibitemOpen
  \bibfield  {author} {\bibinfo {author} {\bibfnamefont {K.}~\bibnamefont {Becker}}, \bibinfo {author} {\bibfnamefont {M.}~\bibnamefont {Becker}},\ and\ \bibinfo {author} {\bibfnamefont {J.~H.}\ \bibnamefont {Schwarz}},\ }\href@noop {} {\emph {\bibinfo {title} {String theory and M-theory: A modern introduction}}}\ (\bibinfo  {publisher} {Cambridge university press},\ \bibinfo {year} {2006})\BibitemShut {NoStop}%
\bibitem [{\citenamefont {Zwiebach}(2004)}]{zwiebach2004}%
  \BibitemOpen
  \bibfield  {author} {\bibinfo {author} {\bibfnamefont {B.}~\bibnamefont {Zwiebach}},\ }\href@noop {} {\emph {\bibinfo {title} {A first course in string theory}}}\ (\bibinfo  {publisher} {Cambridge university press},\ \bibinfo {year} {2004})\BibitemShut {NoStop}%
\bibitem [{\citenamefont {Heisenberg}(1927)}]{heisenberg1927}%
  \BibitemOpen
  \bibfield  {author} {\bibinfo {author} {\bibfnamefont {W.}~\bibnamefont {Heisenberg}},\ }\href@noop {} {\bibfield  {journal} {\bibinfo  {journal} {Zeitschrift f{\"u}r Physik}\ }\textbf {\bibinfo {volume} {43}},\ \bibinfo {pages} {172} (\bibinfo {year} {1927})}\BibitemShut {NoStop}%
\bibitem [{\citenamefont {Kempf}\ \emph {et~al.}(1995)\citenamefont {Kempf}, \citenamefont {Mangano},\ and\ \citenamefont {Mann}}]{kempf1995}%
  \BibitemOpen
  \bibfield  {author} {\bibinfo {author} {\bibfnamefont {A.}~\bibnamefont {Kempf}}, \bibinfo {author} {\bibfnamefont {G.}~\bibnamefont {Mangano}},\ and\ \bibinfo {author} {\bibfnamefont {R.~B.}\ \bibnamefont {Mann}},\ }\href@noop {} {\bibfield  {journal} {\bibinfo  {journal} {Physical Review D}\ }\textbf {\bibinfo {volume} {52}},\ \bibinfo {pages} {1108} (\bibinfo {year} {1995})}\BibitemShut {NoStop}%
\bibitem [{\citenamefont {Shah}\ \emph {et~al.}(2023)\citenamefont {Shah}, \citenamefont {Shaikh}, \citenamefont {Yamin}, \citenamefont {Sahoo}, \citenamefont {Bhat}, \citenamefont {Lone}, \citenamefont {Faizal},\ and\ \citenamefont {Ahsan}}]{shah2023}%
  \BibitemOpen
  \bibfield  {author} {\bibinfo {author} {\bibfnamefont {N.~A.}\ \bibnamefont {Shah}}, \bibinfo {author} {\bibfnamefont {A.}~\bibnamefont {Shaikh}}, \bibinfo {author} {\bibfnamefont {Y.}~\bibnamefont {Yamin}}, \bibinfo {author} {\bibfnamefont {P.}~\bibnamefont {Sahoo}}, \bibinfo {author} {\bibfnamefont {A.}~\bibnamefont {Bhat}}, \bibinfo {author} {\bibfnamefont {S.~A.}\ \bibnamefont {Lone}}, \bibinfo {author} {\bibfnamefont {M.}~\bibnamefont {Faizal}},\ and\ \bibinfo {author} {\bibfnamefont {M.}~\bibnamefont {Ahsan}},\ }\href@noop {} {\bibfield  {journal} {\bibinfo  {journal} {arXiv preprint arXiv:2302.12572}\ } (\bibinfo {year} {2023})}\BibitemShut {NoStop}%
\bibitem [{\citenamefont {Mas{\l}owski}\ \emph {et~al.}(2012)\citenamefont {Mas{\l}owski}, \citenamefont {Nowicki},\ and\ \citenamefont {Tkachuk}}]{maslowski2012}%
  \BibitemOpen
  \bibfield  {author} {\bibinfo {author} {\bibfnamefont {T.}~\bibnamefont {Mas{\l}owski}}, \bibinfo {author} {\bibfnamefont {A.}~\bibnamefont {Nowicki}},\ and\ \bibinfo {author} {\bibfnamefont {V.~M.}\ \bibnamefont {Tkachuk}},\ }\href@noop {} {\bibfield  {journal} {\bibinfo  {journal} {Journal of Physics A: Mathematical and Theoretical}\ }\textbf {\bibinfo {volume} {45}},\ \bibinfo {pages} {075309} (\bibinfo {year} {2012})}\BibitemShut {NoStop}%
\bibitem [{\citenamefont {Moyal}(1949)}]{moyal1949}%
  \BibitemOpen
  \bibfield  {author} {\bibinfo {author} {\bibfnamefont {J.~E.}\ \bibnamefont {Moyal}},\ }in\ \href@noop {} {\emph {\bibinfo {booktitle} {Mathematical Proceedings of the Cambridge Philosophical Society}}},\ Vol.~\bibinfo {volume} {45}\ (\bibinfo {organization} {Cambridge University Press},\ \bibinfo {year} {1949})\ pp.\ \bibinfo {pages} {99--124}\BibitemShut {NoStop}%
\bibitem [{\citenamefont {Sternheimer}(1998)}]{sternheimer1998}%
  \BibitemOpen
  \bibfield  {author} {\bibinfo {author} {\bibfnamefont {D.}~\bibnamefont {Sternheimer}},\ }in\ \href@noop {} {\emph {\bibinfo {booktitle} {AIP Conference Proceedings}}},\ Vol.\ \bibinfo {volume} {453}\ (\bibinfo {organization} {American Institute of Physics},\ \bibinfo {year} {1998})\ pp.\ \bibinfo {pages} {107--145}\BibitemShut {NoStop}%
\bibitem [{\citenamefont {Dirac}(1981)}]{dirac1981}%
  \BibitemOpen
  \bibfield  {author} {\bibinfo {author} {\bibfnamefont {P.~A.~M.}\ \bibnamefont {Dirac}},\ }\href@noop {} {\emph {\bibinfo {title} {The principles of quantum mechanics}}},\ \bibinfo {number} {27}\ (\bibinfo  {publisher} {Oxford university press},\ \bibinfo {year} {1981})\BibitemShut {NoStop}%
\bibitem [{\citenamefont {Ashtekar}\ and\ \citenamefont {Tate}(1994)}]{ashtekar1994}%
  \BibitemOpen
  \bibfield  {author} {\bibinfo {author} {\bibfnamefont {A.}~\bibnamefont {Ashtekar}}\ and\ \bibinfo {author} {\bibfnamefont {R.~S.}\ \bibnamefont {Tate}},\ }\href@noop {} {\bibfield  {journal} {\bibinfo  {journal} {Journal of Mathematical Physics}\ }\textbf {\bibinfo {volume} {35}},\ \bibinfo {pages} {6434} (\bibinfo {year} {1994})}\BibitemShut {NoStop}%
\bibitem [{\citenamefont {Groenewold}(1946)}]{groenewold1946}%
  \BibitemOpen
  \bibfield  {author} {\bibinfo {author} {\bibfnamefont {H.~J.}\ \bibnamefont {Groenewold}},\ }\href@noop {} {\emph {\bibinfo {title} {On the principles of elementary quantum mechanics}}}\ (\bibinfo  {publisher} {Springer},\ \bibinfo {year} {1946})\BibitemShut {NoStop}%
\bibitem [{\citenamefont {Ali}\ and\ \citenamefont {Engli{\v{s}}}(2005)}]{ali2005}%
  \BibitemOpen
  \bibfield  {author} {\bibinfo {author} {\bibfnamefont {S.~T.}\ \bibnamefont {Ali}}\ and\ \bibinfo {author} {\bibfnamefont {M.}~\bibnamefont {Engli{\v{s}}}},\ }\href@noop {} {\bibfield  {journal} {\bibinfo  {journal} {Reviews in Mathematical Physics}\ }\textbf {\bibinfo {volume} {17}},\ \bibinfo {pages} {391} (\bibinfo {year} {2005})}\BibitemShut {NoStop}%
\bibitem [{\citenamefont {Wigner}(1932)}]{wigner1932}%
  \BibitemOpen
  \bibfield  {author} {\bibinfo {author} {\bibfnamefont {E.}~\bibnamefont {Wigner}},\ }\href@noop {} {\bibfield  {journal} {\bibinfo  {journal} {Physical review}\ }\textbf {\bibinfo {volume} {40}},\ \bibinfo {pages} {749} (\bibinfo {year} {1932})}\BibitemShut {NoStop}%
\bibitem [{\citenamefont {Weyl}(1927)}]{weyl1927}%
  \BibitemOpen
  \bibfield  {author} {\bibinfo {author} {\bibfnamefont {H.}~\bibnamefont {Weyl}},\ }\href@noop {} {\bibfield  {journal} {\bibinfo  {journal} {Zeitschrift f{\"u}r Physik}\ }\textbf {\bibinfo {volume} {46}},\ \bibinfo {pages} {1} (\bibinfo {year} {1927})}\BibitemShut {NoStop}%
\bibitem [{\citenamefont {DeWitt}(1957)}]{dewitt1957}%
  \BibitemOpen
  \bibfield  {author} {\bibinfo {author} {\bibfnamefont {B.~S.}\ \bibnamefont {DeWitt}},\ }\href@noop {} {\bibfield  {journal} {\bibinfo  {journal} {Reviews of Modern Physics}\ }\textbf {\bibinfo {volume} {29}},\ \bibinfo {pages} {377} (\bibinfo {year} {1957})}\BibitemShut {NoStop}%
\bibitem [{\citenamefont {Weiss}(1978)}]{weiss1978}%
  \BibitemOpen
  \bibfield  {author} {\bibinfo {author} {\bibfnamefont {U.}~\bibnamefont {Weiss}},\ }\href@noop {} {\bibfield  {journal} {\bibinfo  {journal} {Zeitschrift für Physik B Condensed Matter}\ } (\bibinfo {year} {1978})}\BibitemShut {NoStop}%
\bibitem [{\citenamefont {McCoy}(1932)}]{mccoy1932}%
  \BibitemOpen
  \bibfield  {author} {\bibinfo {author} {\bibfnamefont {N.~H.}\ \bibnamefont {McCoy}},\ }\href@noop {} {\bibfield  {journal} {\bibinfo  {journal} {Proceedings of the National Academy of Sciences}\ }\textbf {\bibinfo {volume} {18}},\ \bibinfo {pages} {674} (\bibinfo {year} {1932})}\BibitemShut {NoStop}%
\bibitem [{\citenamefont {Oppenheim}(2023)}]{oppenheim2023}%
  \BibitemOpen
  \bibfield  {author} {\bibinfo {author} {\bibfnamefont {J.}~\bibnamefont {Oppenheim}},\ }\href@noop {} {\bibfield  {journal} {\bibinfo  {journal} {Physical Review X}\ }\textbf {\bibinfo {volume} {13}},\ \bibinfo {pages} {041040} (\bibinfo {year} {2023})}\BibitemShut {NoStop}%
\bibitem [{\citenamefont {Kafri}\ \emph {et~al.}(2014)\citenamefont {Kafri}, \citenamefont {Taylor},\ and\ \citenamefont {Milburn}}]{kafri2014}%
  \BibitemOpen
  \bibfield  {author} {\bibinfo {author} {\bibfnamefont {D.}~\bibnamefont {Kafri}}, \bibinfo {author} {\bibfnamefont {J.~M.}\ \bibnamefont {Taylor}},\ and\ \bibinfo {author} {\bibfnamefont {G.~J.}\ \bibnamefont {Milburn}},\ }\href@noop {} {\bibfield  {journal} {\bibinfo  {journal} {New Journal of Physics}\ }\textbf {\bibinfo {volume} {16}},\ \bibinfo {pages} {065020} (\bibinfo {year} {2014})}\BibitemShut {NoStop}%
\bibitem [{\citenamefont {Peres}\ and\ \citenamefont {Terno}(2004)}]{peres2004}%
  \BibitemOpen
  \bibfield  {author} {\bibinfo {author} {\bibfnamefont {A.}~\bibnamefont {Peres}}\ and\ \bibinfo {author} {\bibfnamefont {D.~R.}\ \bibnamefont {Terno}},\ }\href@noop {} {\bibfield  {journal} {\bibinfo  {journal} {Reviews of Modern Physics}\ }\textbf {\bibinfo {volume} {76}},\ \bibinfo {pages} {93} (\bibinfo {year} {2004})}\BibitemShut {NoStop}%
\bibitem [{\citenamefont {Altamirano}\ \emph {et~al.}(2018)\citenamefont {Altamirano}, \citenamefont {Corona-Ugalde}, \citenamefont {Mann},\ and\ \citenamefont {Zych}}]{altamirano2018}%
  \BibitemOpen
  \bibfield  {author} {\bibinfo {author} {\bibfnamefont {N.}~\bibnamefont {Altamirano}}, \bibinfo {author} {\bibfnamefont {P.}~\bibnamefont {Corona-Ugalde}}, \bibinfo {author} {\bibfnamefont {R.~B.}\ \bibnamefont {Mann}},\ and\ \bibinfo {author} {\bibfnamefont {M.}~\bibnamefont {Zych}},\ }\href@noop {} {\bibfield  {journal} {\bibinfo  {journal} {Classical and Quantum Gravity}\ }\textbf {\bibinfo {volume} {35}},\ \bibinfo {pages} {145005} (\bibinfo {year} {2018})}\BibitemShut {NoStop}%
\bibitem [{\citenamefont {Becattini}\ and\ \citenamefont {Grossi}(2015)}]{becattini2015}%
  \BibitemOpen
  \bibfield  {author} {\bibinfo {author} {\bibfnamefont {F.}~\bibnamefont {Becattini}}\ and\ \bibinfo {author} {\bibfnamefont {E.}~\bibnamefont {Grossi}},\ }\href@noop {} {\bibfield  {journal} {\bibinfo  {journal} {Physical Review D}\ }\textbf {\bibinfo {volume} {92}},\ \bibinfo {pages} {045037} (\bibinfo {year} {2015})}\BibitemShut {NoStop}%
\bibitem [{\citenamefont {Zubarev}\ \emph {et~al.}(1979)\citenamefont {Zubarev}, \citenamefont {Prozorkevich},\ and\ \citenamefont {Smolyansky}}]{zubarev1979}%
  \BibitemOpen
  \bibfield  {author} {\bibinfo {author} {\bibfnamefont {D.~N.}\ \bibnamefont {Zubarev}}, \bibinfo {author} {\bibfnamefont {A.~V.}\ \bibnamefont {Prozorkevich}},\ and\ \bibinfo {author} {\bibfnamefont {S.~A.}\ \bibnamefont {Smolyansky}},\ }\href@noop {} {\bibfield  {journal} {\bibinfo  {journal} {Teoreticheskaya i Matematicheskaya Fizika}\ }\textbf {\bibinfo {volume} {40}},\ \bibinfo {pages} {394} (\bibinfo {year} {1979})}\BibitemShut {NoStop}%
\bibitem [{\citenamefont {Van~Weert}(1982)}]{van1982}%
  \BibitemOpen
  \bibfield  {author} {\bibinfo {author} {\bibfnamefont {C.~G.}\ \bibnamefont {Van~Weert}},\ }\href@noop {} {\bibfield  {journal} {\bibinfo  {journal} {Annals of Physics}\ }\textbf {\bibinfo {volume} {140}},\ \bibinfo {pages} {133} (\bibinfo {year} {1982})}\BibitemShut {NoStop}%
\bibitem [{\citenamefont {Frolov}\ and\ \citenamefont {Zel’nikov}(1987)}]{frolov1987}%
  \BibitemOpen
  \bibfield  {author} {\bibinfo {author} {\bibfnamefont {V.~P.}\ \bibnamefont {Frolov}}\ and\ \bibinfo {author} {\bibfnamefont {A.~I.}\ \bibnamefont {Zel’nikov}},\ }\href@noop {} {\bibfield  {journal} {\bibinfo  {journal} {Physical Review D}\ }\textbf {\bibinfo {volume} {35}},\ \bibinfo {pages} {3031} (\bibinfo {year} {1987})}\BibitemShut {NoStop}%
\bibitem [{\citenamefont {Zych}\ \emph {et~al.}(2011)\citenamefont {Zych}, \citenamefont {Costa}, \citenamefont {Pikovski},\ and\ \citenamefont {Brukner}}]{zych2011quantum}%
  \BibitemOpen
  \bibfield  {author} {\bibinfo {author} {\bibfnamefont {M.}~\bibnamefont {Zych}}, \bibinfo {author} {\bibfnamefont {F.}~\bibnamefont {Costa}}, \bibinfo {author} {\bibfnamefont {I.}~\bibnamefont {Pikovski}},\ and\ \bibinfo {author} {\bibfnamefont {{\v{C}}.}~\bibnamefont {Brukner}},\ }\href@noop {} {\bibfield  {journal} {\bibinfo  {journal} {Nature communications}\ }\textbf {\bibinfo {volume} {2}},\ \bibinfo {pages} {505} (\bibinfo {year} {2011})}\BibitemShut {NoStop}%
\bibitem [{\citenamefont {Pikovski}\ \emph {et~al.}(2015)\citenamefont {Pikovski}, \citenamefont {Zych}, \citenamefont {Costa},\ and\ \citenamefont {Brukner}}]{pikovski2015universal}%
  \BibitemOpen
  \bibfield  {author} {\bibinfo {author} {\bibfnamefont {I.}~\bibnamefont {Pikovski}}, \bibinfo {author} {\bibfnamefont {M.}~\bibnamefont {Zych}}, \bibinfo {author} {\bibfnamefont {F.}~\bibnamefont {Costa}},\ and\ \bibinfo {author} {\bibfnamefont {{\v{C}}.}~\bibnamefont {Brukner}},\ }\href@noop {} {\bibfield  {journal} {\bibinfo  {journal} {Nature Physics}\ }\textbf {\bibinfo {volume} {11}},\ \bibinfo {pages} {668} (\bibinfo {year} {2015})}\BibitemShut {NoStop}%
\bibitem [{\citenamefont {Pikovski}\ \emph {et~al.}(2017)\citenamefont {Pikovski}, \citenamefont {Zych}, \citenamefont {Costa},\ and\ \citenamefont {Brukner}}]{pikovski2017time}%
  \BibitemOpen
  \bibfield  {author} {\bibinfo {author} {\bibfnamefont {I.}~\bibnamefont {Pikovski}}, \bibinfo {author} {\bibfnamefont {M.}~\bibnamefont {Zych}}, \bibinfo {author} {\bibfnamefont {F.}~\bibnamefont {Costa}},\ and\ \bibinfo {author} {\bibfnamefont {{\v{C}}.}~\bibnamefont {Brukner}},\ }\href@noop {} {\bibfield  {journal} {\bibinfo  {journal} {New Journal of Physics}\ }\textbf {\bibinfo {volume} {19}},\ \bibinfo {pages} {025011} (\bibinfo {year} {2017})}\BibitemShut {NoStop}%
\bibitem [{\citenamefont {Paczos}\ \emph {et~al.}(2024)\citenamefont {Paczos}, \citenamefont {D{\k{e}}bski}, \citenamefont {Grochowski}, \citenamefont {Smith},\ and\ \citenamefont {Dragan}}]{paczos2024quantum}%
  \BibitemOpen
  \bibfield  {author} {\bibinfo {author} {\bibfnamefont {J.}~\bibnamefont {Paczos}}, \bibinfo {author} {\bibfnamefont {K.}~\bibnamefont {D{\k{e}}bski}}, \bibinfo {author} {\bibfnamefont {P.~T.}\ \bibnamefont {Grochowski}}, \bibinfo {author} {\bibfnamefont {A.~R.}\ \bibnamefont {Smith}},\ and\ \bibinfo {author} {\bibfnamefont {A.}~\bibnamefont {Dragan}},\ }\href@noop {} {\bibfield  {journal} {\bibinfo  {journal} {Quantum}\ }\textbf {\bibinfo {volume} {8}},\ \bibinfo {pages} {1338} (\bibinfo {year} {2024})}\BibitemShut {NoStop}%
\bibitem [{\citenamefont {Cepollaro}\ \emph {et~al.}(2023)\citenamefont {Cepollaro}, \citenamefont {Giacomini},\ and\ \citenamefont {Paris}}]{cepollaro2023gravitational}%
  \BibitemOpen
  \bibfield  {author} {\bibinfo {author} {\bibfnamefont {C.}~\bibnamefont {Cepollaro}}, \bibinfo {author} {\bibfnamefont {F.}~\bibnamefont {Giacomini}},\ and\ \bibinfo {author} {\bibfnamefont {M.~G.}\ \bibnamefont {Paris}},\ }\href@noop {} {\bibfield  {journal} {\bibinfo  {journal} {Quantum}\ }\textbf {\bibinfo {volume} {7}},\ \bibinfo {pages} {946} (\bibinfo {year} {2023})}\BibitemShut {NoStop}%
\bibitem [{\citenamefont {Roura}\ \emph {et~al.}(2021)\citenamefont {Roura}, \citenamefont {Schubert}, \citenamefont {Schlippert},\ and\ \citenamefont {Rasel}}]{roura2021measuring}%
  \BibitemOpen
  \bibfield  {author} {\bibinfo {author} {\bibfnamefont {A.}~\bibnamefont {Roura}}, \bibinfo {author} {\bibfnamefont {C.}~\bibnamefont {Schubert}}, \bibinfo {author} {\bibfnamefont {D.}~\bibnamefont {Schlippert}},\ and\ \bibinfo {author} {\bibfnamefont {E.~M.}\ \bibnamefont {Rasel}},\ }\href@noop {} {\bibfield  {journal} {\bibinfo  {journal} {Physical Review D}\ }\textbf {\bibinfo {volume} {104}},\ \bibinfo {pages} {084001} (\bibinfo {year} {2021})}\BibitemShut {NoStop}%
\bibitem [{\citenamefont {Smith}\ and\ \citenamefont {Ahmadi}(2020)}]{smith2020quantum}%
  \BibitemOpen
  \bibfield  {author} {\bibinfo {author} {\bibfnamefont {A.~R.}\ \bibnamefont {Smith}}\ and\ \bibinfo {author} {\bibfnamefont {M.}~\bibnamefont {Ahmadi}},\ }\href@noop {} {\bibfield  {journal} {\bibinfo  {journal} {Nature communications}\ }\textbf {\bibinfo {volume} {11}},\ \bibinfo {pages} {5360} (\bibinfo {year} {2020})}\BibitemShut {NoStop}%
\bibitem [{\citenamefont {D{\k{e}}bski}\ \emph {et~al.}(2024)\citenamefont {D{\k{e}}bski}, \citenamefont {Grochowski}, \citenamefont {Demkowicz-Dobrza{\'n}ski},\ and\ \citenamefont {Dragan}}]{dkebski2024universality}%
  \BibitemOpen
  \bibfield  {author} {\bibinfo {author} {\bibfnamefont {K.}~\bibnamefont {D{\k{e}}bski}}, \bibinfo {author} {\bibfnamefont {P.~T.}\ \bibnamefont {Grochowski}}, \bibinfo {author} {\bibfnamefont {R.}~\bibnamefont {Demkowicz-Dobrza{\'n}ski}},\ and\ \bibinfo {author} {\bibfnamefont {A.}~\bibnamefont {Dragan}},\ }\href@noop {} {\bibfield  {journal} {\bibinfo  {journal} {Classical and Quantum Gravity}\ }\textbf {\bibinfo {volume} {41}},\ \bibinfo {pages} {135014} (\bibinfo {year} {2024})}\BibitemShut {NoStop}%
\bibitem [{\citenamefont {Zych}(2017)}]{zych2017quantum}%
  \BibitemOpen
  \bibfield  {author} {\bibinfo {author} {\bibfnamefont {M.}~\bibnamefont {Zych}},\ }\href@noop {} {\emph {\bibinfo {title} {Quantum systems under gravitational time dilation}}}\ (\bibinfo  {publisher} {Springer},\ \bibinfo {year} {2017})\BibitemShut {NoStop}%
\bibitem [{\citenamefont {Bose}\ \emph {et~al.}(2017)\citenamefont {Bose}, \citenamefont {Mazumdar}, \citenamefont {Morley}, \citenamefont {Ulbricht}, \citenamefont {Toro{\v{s}}}, \citenamefont {Paternostro}, \citenamefont {Geraci}, \citenamefont {Barker}, \citenamefont {Kim},\ and\ \citenamefont {Milburn}}]{bose2017}%
  \BibitemOpen
  \bibfield  {author} {\bibinfo {author} {\bibfnamefont {S.}~\bibnamefont {Bose}}, \bibinfo {author} {\bibfnamefont {A.}~\bibnamefont {Mazumdar}}, \bibinfo {author} {\bibfnamefont {G.~W.}\ \bibnamefont {Morley}}, \bibinfo {author} {\bibfnamefont {H.}~\bibnamefont {Ulbricht}}, \bibinfo {author} {\bibfnamefont {M.}~\bibnamefont {Toro{\v{s}}}}, \bibinfo {author} {\bibfnamefont {M.}~\bibnamefont {Paternostro}}, \bibinfo {author} {\bibfnamefont {A.~A.}\ \bibnamefont {Geraci}}, \bibinfo {author} {\bibfnamefont {P.~F.}\ \bibnamefont {Barker}}, \bibinfo {author} {\bibfnamefont {M.}~\bibnamefont {Kim}},\ and\ \bibinfo {author} {\bibfnamefont {G.}~\bibnamefont {Milburn}},\ }\href@noop {} {\bibfield  {journal} {\bibinfo  {journal} {Physical review letters}\ }\textbf {\bibinfo {volume} {119}},\ \bibinfo {pages} {240401} (\bibinfo {year} {2017})}\BibitemShut {NoStop}%
\bibitem [{\citenamefont {Marletto}\ and\ \citenamefont {Vedral}(2017)}]{marletto2017}%
  \BibitemOpen
  \bibfield  {author} {\bibinfo {author} {\bibfnamefont {C.}~\bibnamefont {Marletto}}\ and\ \bibinfo {author} {\bibfnamefont {V.}~\bibnamefont {Vedral}},\ }\href@noop {} {\bibfield  {journal} {\bibinfo  {journal} {Physical review letters}\ }\textbf {\bibinfo {volume} {119}},\ \bibinfo {pages} {240402} (\bibinfo {year} {2017})}\BibitemShut {NoStop}%
\bibitem [{\citenamefont {Mart{\'\i}n-Mart{\'\i}nez}\ and\ \citenamefont {Perche}(2023)}]{martin2023}%
  \BibitemOpen
  \bibfield  {author} {\bibinfo {author} {\bibfnamefont {E.}~\bibnamefont {Mart{\'\i}n-Mart{\'\i}nez}}\ and\ \bibinfo {author} {\bibfnamefont {T.~R.}\ \bibnamefont {Perche}},\ }\href@noop {} {\bibfield  {journal} {\bibinfo  {journal} {Physical Review D}\ }\textbf {\bibinfo {volume} {108}},\ \bibinfo {pages} {L101702} (\bibinfo {year} {2023})}\BibitemShut {NoStop}%
\bibitem [{\citenamefont {Perche}\ and\ \citenamefont {Mart{\'\i}n-Mart{\'\i}nez}(2023)}]{perche2023}%
  \BibitemOpen
  \bibfield  {author} {\bibinfo {author} {\bibfnamefont {R.~T.}\ \bibnamefont {Perche}}\ and\ \bibinfo {author} {\bibfnamefont {E.}~\bibnamefont {Mart{\'\i}n-Mart{\'\i}nez}},\ }\href@noop {} {\bibfield  {journal} {\bibinfo  {journal} {Physical Review A}\ }\textbf {\bibinfo {volume} {107}},\ \bibinfo {pages} {042612} (\bibinfo {year} {2023})}\BibitemShut {NoStop}%
\bibitem [{\citenamefont {Tobar}\ \emph {et~al.}(2024)\citenamefont {Tobar}, \citenamefont {Manikandan}, \citenamefont {Beitel},\ and\ \citenamefont {Pikovski}}]{tobar2024detecting}%
  \BibitemOpen
  \bibfield  {author} {\bibinfo {author} {\bibfnamefont {G.}~\bibnamefont {Tobar}}, \bibinfo {author} {\bibfnamefont {S.~K.}\ \bibnamefont {Manikandan}}, \bibinfo {author} {\bibfnamefont {T.}~\bibnamefont {Beitel}},\ and\ \bibinfo {author} {\bibfnamefont {I.}~\bibnamefont {Pikovski}},\ }\href@noop {} {\bibfield  {journal} {\bibinfo  {journal} {Nature Communications}\ }\textbf {\bibinfo {volume} {15}},\ \bibinfo {pages} {7229} (\bibinfo {year} {2024})}\BibitemShut {NoStop}%
\bibitem [{\citenamefont {Suzuki}\ \emph {et~al.}(1980)\citenamefont {Suzuki}, \citenamefont {Hirshfeld},\ and\ \citenamefont {Leschke}}]{suzuki1980}%
  \BibitemOpen
  \bibfield  {author} {\bibinfo {author} {\bibfnamefont {T.}~\bibnamefont {Suzuki}}, \bibinfo {author} {\bibfnamefont {A.}~\bibnamefont {Hirshfeld}},\ and\ \bibinfo {author} {\bibfnamefont {H.}~\bibnamefont {Leschke}},\ }\href@noop {} {\bibfield  {journal} {\bibinfo  {journal} {Progress of Theoretical Physics}\ }\textbf {\bibinfo {volume} {63}},\ \bibinfo {pages} {287} (\bibinfo {year} {1980})}\BibitemShut {NoStop}%
\bibitem [{\citenamefont {Christodoulakis}\ and\ \citenamefont {Zanelli}(1986)}]{christodoulakis1986}%
  \BibitemOpen
  \bibfield  {author} {\bibinfo {author} {\bibfnamefont {T.}~\bibnamefont {Christodoulakis}}\ and\ \bibinfo {author} {\bibfnamefont {J.}~\bibnamefont {Zanelli}},\ }\href@noop {} {\bibfield  {journal} {\bibinfo  {journal} {Il Nuovo Cimento B (1971-1996)}\ }\textbf {\bibinfo {volume} {93}},\ \bibinfo {pages} {1} (\bibinfo {year} {1986})}\BibitemShut {NoStop}%
\bibitem [{\citenamefont {Anderson}(2012)}]{anderson2012}%
  \BibitemOpen
  \bibfield  {author} {\bibinfo {author} {\bibfnamefont {E.}~\bibnamefont {Anderson}},\ }\href@noop {} {\bibfield  {journal} {\bibinfo  {journal} {Annalen der Physik}\ }\textbf {\bibinfo {volume} {524}},\ \bibinfo {pages} {757} (\bibinfo {year} {2012})}\BibitemShut {NoStop}%
\bibitem [{\citenamefont {Khandelwal}\ \emph {et~al.}(2020)\citenamefont {Khandelwal}, \citenamefont {Lock},\ and\ \citenamefont {Woods}}]{khandelwal2020}%
  \BibitemOpen
  \bibfield  {author} {\bibinfo {author} {\bibfnamefont {S.}~\bibnamefont {Khandelwal}}, \bibinfo {author} {\bibfnamefont {M.~P.}\ \bibnamefont {Lock}},\ and\ \bibinfo {author} {\bibfnamefont {M.~P.}\ \bibnamefont {Woods}},\ }\href@noop {} {\bibfield  {journal} {\bibinfo  {journal} {Quantum}\ }\textbf {\bibinfo {volume} {4}},\ \bibinfo {pages} {309} (\bibinfo {year} {2020})}\BibitemShut {NoStop}%
\bibitem [{\citenamefont {D{\k{e}}bski}\ \emph {et~al.}(2022)\citenamefont {D{\k{e}}bski}, \citenamefont {Grochowski}, \citenamefont {Demkowicz-Dobrzanski},\ and\ \citenamefont {Dragan}}]{debski2022}%
  \BibitemOpen
  \bibfield  {author} {\bibinfo {author} {\bibfnamefont {K.}~\bibnamefont {D{\k{e}}bski}}, \bibinfo {author} {\bibfnamefont {P.~T.}\ \bibnamefont {Grochowski}}, \bibinfo {author} {\bibfnamefont {R.}~\bibnamefont {Demkowicz-Dobrzanski}},\ and\ \bibinfo {author} {\bibfnamefont {A.}~\bibnamefont {Dragan}},\ }\href@noop {} {\bibfield  {journal} {\bibinfo  {journal} {Classical and Quantum Gravity}\ } (\bibinfo {year} {2022})}\BibitemShut {NoStop}%
\bibitem [{\citenamefont {Einstein}(1915)}]{einstein1915}%
  \BibitemOpen
  \bibfield  {author} {\bibinfo {author} {\bibfnamefont {A.}~\bibnamefont {Einstein}},\ }\href@noop {} {\bibfield  {journal} {\bibinfo  {journal} {Sitzungsberichte der K{\"o}niglich Preu{\ss}ischen Akademie der Wissenschaften}\ ,\ \bibinfo {pages} {844}} (\bibinfo {year} {1915})}\BibitemShut {NoStop}%
\bibitem [{\citenamefont {Marto}(2021)}]{marto2021}%
  \BibitemOpen
  \bibfield  {author} {\bibinfo {author} {\bibfnamefont {J.}~\bibnamefont {Marto}},\ }\href@noop {} {\bibfield  {journal} {\bibinfo  {journal} {Universe}\ }\textbf {\bibinfo {volume} {7}},\ \bibinfo {pages} {297} (\bibinfo {year} {2021})}\BibitemShut {NoStop}%
\bibitem [{\citenamefont {Tolman}(1928)}]{tolman1928}%
  \BibitemOpen
  \bibfield  {author} {\bibinfo {author} {\bibfnamefont {R.~C.}\ \bibnamefont {Tolman}},\ }\href@noop {} {\bibfield  {journal} {\bibinfo  {journal} {Proceedings of the National Academy of Sciences}\ }\textbf {\bibinfo {volume} {14}},\ \bibinfo {pages} {701} (\bibinfo {year} {1928})}\BibitemShut {NoStop}%
\bibitem [{\citenamefont {Tolman}\ and\ \citenamefont {Ehrenfest}(1930)}]{tolman1930}%
  \BibitemOpen
  \bibfield  {author} {\bibinfo {author} {\bibfnamefont {R.~C.}\ \bibnamefont {Tolman}}\ and\ \bibinfo {author} {\bibfnamefont {P.}~\bibnamefont {Ehrenfest}},\ }\href@noop {} {\bibfield  {journal} {\bibinfo  {journal} {Physical Review}\ }\textbf {\bibinfo {volume} {36}},\ \bibinfo {pages} {1791} (\bibinfo {year} {1930})}\BibitemShut {NoStop}%
\bibitem [{\citenamefont {Vall{\'e}e}\ and\ \citenamefont {Soares}(2010)}]{vallee2010airy}%
  \BibitemOpen
  \bibfield  {author} {\bibinfo {author} {\bibfnamefont {O.}~\bibnamefont {Vall{\'e}e}}\ and\ \bibinfo {author} {\bibfnamefont {M.}~\bibnamefont {Soares}},\ }\href@noop {} {\emph {\bibinfo {title} {Airy functions and applications to physics}}}\ (\bibinfo  {publisher} {World Scientific Publishing Company},\ \bibinfo {year} {2010})\BibitemShut {NoStop}%
\bibitem [{\citenamefont {Bieniek}(1977)}]{bieniek1977uniform}%
  \BibitemOpen
  \bibfield  {author} {\bibinfo {author} {\bibfnamefont {R.}~\bibnamefont {Bieniek}},\ }\href@noop {} {\bibfield  {journal} {\bibinfo  {journal} {Physical Review A}\ }\textbf {\bibinfo {volume} {15}},\ \bibinfo {pages} {1513} (\bibinfo {year} {1977})}\BibitemShut {NoStop}%
\bibitem [{\citenamefont {Tong}(2012)}]{tong_statistical_physics}%
  \BibitemOpen
  \bibfield  {author} {\bibinfo {author} {\bibfnamefont {D.}~\bibnamefont {Tong}},\ }\href@noop {} {\bibinfo {title} {Statistical physics}},\ \bibinfo {howpublished} {University of Cambridge} (\bibinfo {year} {2012})\BibitemShut {NoStop}%
\bibitem [{\citenamefont {Rouabhia}\ \emph {et~al.}(2023)\citenamefont {Rouabhia}, \citenamefont {Boumali},\ and\ \citenamefont {Hassanabadi}}]{rouabhia2023effect}%
  \BibitemOpen
  \bibfield  {author} {\bibinfo {author} {\bibfnamefont {T.~I.}\ \bibnamefont {Rouabhia}}, \bibinfo {author} {\bibfnamefont {A.}~\bibnamefont {Boumali}},\ and\ \bibinfo {author} {\bibfnamefont {H.}~\bibnamefont {Hassanabadi}},\ }\href@noop {} {\bibfield  {journal} {\bibinfo  {journal} {Physics of Particles and Nuclei Letters}\ }\textbf {\bibinfo {volume} {20}},\ \bibinfo {pages} {112} (\bibinfo {year} {2023})}\BibitemShut {NoStop}%
\bibitem [{\citenamefont {Deppner}\ \emph {et~al.}(2021)\citenamefont {Deppner}, \citenamefont {Herr}, \citenamefont {Cornelius}, \citenamefont {Stromberger}, \citenamefont {Sternke}, \citenamefont {Grzeschik}, \citenamefont {Grote}, \citenamefont {Rudolph}, \citenamefont {Herrmann}, \citenamefont {Krutzik} \emph {et~al.}}]{deppner2021collective}%
  \BibitemOpen
  \bibfield  {author} {\bibinfo {author} {\bibfnamefont {C.}~\bibnamefont {Deppner}}, \bibinfo {author} {\bibfnamefont {W.}~\bibnamefont {Herr}}, \bibinfo {author} {\bibfnamefont {M.}~\bibnamefont {Cornelius}}, \bibinfo {author} {\bibfnamefont {P.}~\bibnamefont {Stromberger}}, \bibinfo {author} {\bibfnamefont {T.}~\bibnamefont {Sternke}}, \bibinfo {author} {\bibfnamefont {C.}~\bibnamefont {Grzeschik}}, \bibinfo {author} {\bibfnamefont {A.}~\bibnamefont {Grote}}, \bibinfo {author} {\bibfnamefont {J.}~\bibnamefont {Rudolph}}, \bibinfo {author} {\bibfnamefont {S.}~\bibnamefont {Herrmann}}, \bibinfo {author} {\bibfnamefont {M.}~\bibnamefont {Krutzik}}, \emph {et~al.},\ }\href@noop {} {\bibfield  {journal} {\bibinfo  {journal} {Physical Review Letters}\ }\textbf {\bibinfo {volume} {127}},\ \bibinfo {pages} {100401} (\bibinfo {year} {2021})}\BibitemShut {NoStop}%
\bibitem [{\citenamefont {Pandey}\ \emph {et~al.}(2019)\citenamefont {Pandey}, \citenamefont {Mas}, \citenamefont {Drougakis}, \citenamefont {Thekkeppatt}, \citenamefont {Bolpasi}, \citenamefont {Vasilakis}, \citenamefont {Poulios},\ and\ \citenamefont {von Klitzing}}]{pandey2019hypersonic}%
  \BibitemOpen
  \bibfield  {author} {\bibinfo {author} {\bibfnamefont {S.}~\bibnamefont {Pandey}}, \bibinfo {author} {\bibfnamefont {H.}~\bibnamefont {Mas}}, \bibinfo {author} {\bibfnamefont {G.}~\bibnamefont {Drougakis}}, \bibinfo {author} {\bibfnamefont {P.}~\bibnamefont {Thekkeppatt}}, \bibinfo {author} {\bibfnamefont {V.}~\bibnamefont {Bolpasi}}, \bibinfo {author} {\bibfnamefont {G.}~\bibnamefont {Vasilakis}}, \bibinfo {author} {\bibfnamefont {K.}~\bibnamefont {Poulios}},\ and\ \bibinfo {author} {\bibfnamefont {W.}~\bibnamefont {von Klitzing}},\ }\href@noop {} {\bibfield  {journal} {\bibinfo  {journal} {Nature}\ }\textbf {\bibinfo {volume} {570}},\ \bibinfo {pages} {205} (\bibinfo {year} {2019})}\BibitemShut {NoStop}%
\bibitem [{\citenamefont {Wu}\ and\ \citenamefont {Niu}(2003)}]{wu2003superfluidity}%
  \BibitemOpen
  \bibfield  {author} {\bibinfo {author} {\bibfnamefont {B.}~\bibnamefont {Wu}}\ and\ \bibinfo {author} {\bibfnamefont {Q.}~\bibnamefont {Niu}},\ }\href@noop {} {\bibfield  {journal} {\bibinfo  {journal} {New journal of Physics}\ }\textbf {\bibinfo {volume} {5}},\ \bibinfo {pages} {104} (\bibinfo {year} {2003})}\BibitemShut {NoStop}%
\bibitem [{\citenamefont {Choi}\ and\ \citenamefont {Niu}(1999)}]{choi1999bose}%
  \BibitemOpen
  \bibfield  {author} {\bibinfo {author} {\bibfnamefont {D.-I.}\ \bibnamefont {Choi}}\ and\ \bibinfo {author} {\bibfnamefont {Q.}~\bibnamefont {Niu}},\ }\href@noop {} {\bibfield  {journal} {\bibinfo  {journal} {Physical Review Letters}\ }\textbf {\bibinfo {volume} {82}},\ \bibinfo {pages} {2022} (\bibinfo {year} {1999})}\BibitemShut {NoStop}%
\bibitem [{\citenamefont {Burger}\ \emph {et~al.}(2001)\citenamefont {Burger}, \citenamefont {Cataliotti}, \citenamefont {Fort}, \citenamefont {Minardi}, \citenamefont {Inguscio}, \citenamefont {Chiofalo},\ and\ \citenamefont {Tosi}}]{burger2001superfluid}%
  \BibitemOpen
  \bibfield  {author} {\bibinfo {author} {\bibfnamefont {S.}~\bibnamefont {Burger}}, \bibinfo {author} {\bibfnamefont {F.~S.}\ \bibnamefont {Cataliotti}}, \bibinfo {author} {\bibfnamefont {C.}~\bibnamefont {Fort}}, \bibinfo {author} {\bibfnamefont {F.}~\bibnamefont {Minardi}}, \bibinfo {author} {\bibfnamefont {M.}~\bibnamefont {Inguscio}}, \bibinfo {author} {\bibfnamefont {M.~L.}\ \bibnamefont {Chiofalo}},\ and\ \bibinfo {author} {\bibfnamefont {M.}~\bibnamefont {Tosi}},\ }\href@noop {} {\bibfield  {journal} {\bibinfo  {journal} {Physical Review Letters}\ }\textbf {\bibinfo {volume} {86}},\ \bibinfo {pages} {4447} (\bibinfo {year} {2001})}\BibitemShut {NoStop}%
\bibitem [{\citenamefont {Newsome}\ and\ \citenamefont {Andrei}(2004)}]{newsome2004relaxation}%
  \BibitemOpen
  \bibfield  {author} {\bibinfo {author} {\bibfnamefont {R.~W.}\ \bibnamefont {Newsome}}\ and\ \bibinfo {author} {\bibfnamefont {E.~Y.}\ \bibnamefont {Andrei}},\ }\href@noop {} {\bibfield  {journal} {\bibinfo  {journal} {Review of scientific instruments}\ }\textbf {\bibinfo {volume} {75}},\ \bibinfo {pages} {104} (\bibinfo {year} {2004})}\BibitemShut {NoStop}%
\bibitem [{\citenamefont {P{\"o}tting}\ \emph {et~al.}(2001)\citenamefont {P{\"o}tting}, \citenamefont {Cramer}, \citenamefont {Schwalb}, \citenamefont {Pu},\ and\ \citenamefont {Meystre}}]{potting2001coherent}%
  \BibitemOpen
  \bibfield  {author} {\bibinfo {author} {\bibfnamefont {S.}~\bibnamefont {P{\"o}tting}}, \bibinfo {author} {\bibfnamefont {M.}~\bibnamefont {Cramer}}, \bibinfo {author} {\bibfnamefont {C.~H.}\ \bibnamefont {Schwalb}}, \bibinfo {author} {\bibfnamefont {H.}~\bibnamefont {Pu}},\ and\ \bibinfo {author} {\bibfnamefont {P.}~\bibnamefont {Meystre}},\ }\href@noop {} {\bibfield  {journal} {\bibinfo  {journal} {Physical Review A}\ }\textbf {\bibinfo {volume} {64}},\ \bibinfo {pages} {023604} (\bibinfo {year} {2001})}\BibitemShut {NoStop}%
\bibitem [{\citenamefont {Collings}(1986)}]{collings1986low}%
  \BibitemOpen
  \bibfield  {author} {\bibinfo {author} {\bibfnamefont {E.~W.}\ \bibnamefont {Collings}},\ }\href@noop {} {\bibfield  {journal} {\bibinfo  {journal} {Applied Superconductivity, Metallurgy, and Physics of Titanium Alloys: Fundamentals Alloy Superconductors: Their Metallurgical, Physical, and Magnetic-Mixed-State Properties}\ ,\ \bibinfo {pages} {307}} (\bibinfo {year} {1986})}\BibitemShut {NoStop}%
\bibitem [{\citenamefont {Sajnok}(2025)}]{code}%
  \BibitemOpen
  \bibfield  {author} {\bibinfo {author} {\bibfnamefont {K.}~\bibnamefont {Sajnok}},\ }\bibfield  {journal} {\bibinfo  {journal} {Zenodo}\ }\href {https://doi.org/10.5281/zenodo.16384540} {10.5281/zenodo.16384540} (\bibinfo {year} {2025})\BibitemShut {NoStop}%
\end{thebibliography}%

\appendix
\section{Relativistic Quantization Procedure} \label{A}
    Let us consider the classical kinetic component of the Eq.\eqref{eq:energy_classical}:
    \begin{align}
         E_p=\sqrt{m^2c^4 - c^2 g^{ij} p_i p_j},       
    \end{align}
    that we aim to quantize. For simplicity, let's assume that the particle's momentum and position operators align with the radial coordinate. Consequently, the only spacelike component of the metric tensor that multiplies a non-vanishing term in Eq.\eqref{eq:energy_classical} is $g^{rr}$. Given the commutation relation $[\hat{p},\hat{x}]=-\ii\hbar$, there exists ambiguity in the ordering of momentum and position operators. Consider an indeterminate ordering of two momentum operators with a $k$-th power position operator: 
    \begin{align}
        : \hat{x}^k \hat{p}^2 : = \alpha \hat{p}^2 \hat{x}^k + \beta \hat{p} \hat{x}^k \hat{p} + \gamma \hat{x}^k \hat{p}^2,
    \end{align}
    where, to satisfy the classical limit of the commuting scalars $x$ and $p$, the three complex parameters satisfy $\alpha+ \beta+ \gamma =1$. The condition for the operator to remain Hermitian imposes that $\gamma = \alpha ^*$ and $ \beta = \beta^*$. Hence, the position-momentum operator can be expressed as:
    \begin{align} \label{eq:xp2-1a}
        : \hat{x}^k \hat{p}^2 : \;  = \alpha \hat{p}^2 \hat{x}^k + (1-2\ii\operatorname{Im} \alpha) \hat{p} \hat{x}^k \hat{p} + \alpha^* \hat{x}^k \hat{p}^2,
    \end{align}
    for some complex $\alpha$. With the commutation relation $[\hat{p},\hat{x}]=-\ii\hbar$, Eq.\eqref{eq:xp2-1a} becomes:
        \begin{align} \label{eq:xp2-2a}
            : \hat{x}^k \hat{p}^2 : \;  = \hat{x}^k \hat{p}^2 -\ii (1-2\ii\operatorname{Im} \alpha) k \hbar \hat{x}^{k-1} \hat{p} - \alpha k(k-1)\hbar^2 \hat{x}^{k-2}.
        \end{align}
    In determining the operator ordering in Eq.\eqref{eq:xp2-1a}, we have omitted configurations where operators with powers of $\hat{x}$ act on $\hat{p}$ from both sides. The exclusion is justified since the quantization process involves a Taylor series expansion of the metric tensor, as outlined below.

    As the metric component now depends on the position operator, but the metric tensor is defined only for scalar variables, a covariant Taylor series expansion is necessary with $\frac{\hat{x}\phi}{r\phi} \ll 1$:
        \begin{align}
            : g^{rr}(r+\hat{x}_r) \hat{p}_r^2 : \; = g^{rr}(r) \hat{p}_r^2 + \sum_{k=1}^\infty \partial_r^k g^{rr}(r) : \hat{x}_r^k \hat{p}_r^2 :,
        \end{align}
    where $: \hat{x}_r^k \hat{p}_r^2 :$ is given by \eqref{eq:xp2-2}. Hereafter, subscripts $r$ will be omitted for the position and momentum operators.
    
    The kinetic Hamiltonian derived from Eq.\eqref{eq:energy_classical} is:
    \begin{align} \label{eq:h-inv-pa}
        \hat{H}_p = \sqrt{ \hat{H_0}^{2} - c^2 : g^{rr}(r+\hat{x}) \hat{p}^2:}.        
    \end{align}        
    By introducing the time-dilation operator $\sqrt{g_{00}(r+\hat{x})}$, we can express the relativistically invariant Hermitian Hamiltonian of a particle with quantum degrees of freedom as follows:
        \begin{align} \label{eq:h-inv-1a}
            \hat{H} = \gamma^* \sqrt{g_{00}(r+\hat{x})}^* \hat{H}_p + \gamma H_p\sqrt{g_{00}(r+\hat{x})},
        \end{align}
    where $\gamma$ is a complex parameter such that $2 \text{Re}\gamma = 1$. After expanding in a Taylor series to the first order in $\frac{\hat{x}\psi}{r\psi}$ and $\frac{c^2 :g^{rr}(r+\hat{x}) \hat{p}^2 :\,\psi}{\hat{H_0}^{2}\psi}$, Eq.\eqref{eq:h-inv-1a}, with substituted Eq.\eqref{eq:h-inv-pa}, becomes:
        \begin{align} \label{eq:h-inv-2a}
            \hat{H} =  &\hat{H_0} \sqrt{g_{00}} \nonumber \\
            &\Bigg[ \gamma^* \left(1+\frac{g'_{00}}{2g_{00}} \hat{x} \right) \left( 1 - \frac{c^2} {2\hat{H_0}^2} \left( g^{rr}\hat{p}^2 + g'^{rr} :\hat{x} \hat{p}^2: \right) \right) \nonumber\\
            &+\gamma \left( 1 - \frac{c^2} {2\hat{H_0}^2} \left( g^{rr}\hat{p}^2 + g'^{rr} :\hat{x} \hat{p}^2: \right) \right)  \left(1+\frac{g'_{00}}{2g_{00}} \hat{x} \right) \Bigg],
        \end{align}
    where $g^{\mu\nu}(r) \equiv g^{\mu\nu}$ and $\partial_r g^{\mu\nu}(r) \equiv g'^{\mu\nu}$. Calculation of the higher-order terms in the Taylor series expansion may enable the experimental measurement of the real part of the ordering parameter $\alpha$ (from the $\alpha k(k-1) \hbar^2 \hat{x}^{k-2}$ term in Eq.\eqref{eq:xp2-2a}). Nevertheless, we focus exclusively on the first order, as even this proves challenging in practice. Substituting Eq.\eqref{eq:xp2-2a} in the Hamiltonian Eq.\eqref{eq:h-inv-2a} and once again, by utilizing the commutation relation and omitting terms of order $\mathcal{O}(\hat{x}^2/r^2)$ in the small quantum position perturbation limit, we arrive at:
        \begin{align} \label{eq:h-inv-finala}
            \hat{H} =  \hat{H_0} \sqrt{g_{00}} \Bigg[&1+\frac{g'_{00}}{2g_{00}} \hat{x} \nonumber 
            \\&- \frac{ c^2} {2\hat{H_0}^2} \left( g^{rr} + \left(g'^{rr} + \frac{ g'_{00}g^{rr}}{2g_{00}} \right) \hat{x} \right) \hat{p}^2 \nonumber \\
            &+\frac{ \ii \hbar c^2} {2\hat{H_0}^2} \left((1-2\ii\operatorname{Im} \alpha) g'^{rr} +\gamma\frac{ g'_{00}g^{rr}}{g_{00}} \right) \hat{p}  \Bigg].
        \end{align}
    The physical implications of Eq.~\eqref{eq:h-inv-finala} are analyzed in Sec.~\ref{sec:iii}.

\section{Semiclassical Specific Heat in Rindler Frame} \label{B}
    The partition function in quantum regime can be approximated as the integral over the phase space, if the temperature is greater than the specific energy scale of the quantum system. That is exactly, what we assume, writing:
    \begin{align} \label{eq:appb-0}
        Z_1 &= \int \limits_{-\frac{c^2}{g}}^\infty \frac{\dd x}{2\pi \ii \hbar} \int \limits_{-\infty}^\infty \dd{p} \ee^{-\beta mc^2\left(1+\frac{g x}{c^2}\right)\left(1+\frac{p^2}{2m^2c^2}\right) + \ii \beta \gamma \frac{\hbar g p}{mc^2}}.
    \end{align}
    We substitute $\chi = 1 + \frac{gx}{c^2}$ and integrate over $\chi$ from $0$ to $\infty$, obtaining:
    \begin{align} \label{eq:appb-1}
        Z_1 &= \frac{mc^2}{\pi \ii \beta \hbar g} \int \limits_{-\infty}^\infty \dd{p} \frac{ \ee^{\ii \beta \gamma \frac{ \hbar g p}{mc^2}}}{2m^2c^2+p^2},
    \end{align}
    that is an integral with two residual values $p=\pm \ii \sqrt{2}mc$. Integrating over upper semicircle in the complex plane (assuming $\mathcal{I}(\gamma)>0$), we obtain:
    \begin{align} \label{eq:appb-2}
        Z_1 &= \frac{c}{\sqrt{2} \beta \hbar g} \ee^{-\gamma \frac{ \sqrt{2} \beta \hbar g}{c}},
    \end{align}
  
    Next, we use the Boltzmann statistics $Z (N,T)=\frac{Z_1^N(T)}{N!}$, that is valid above the critical temperature \cite{tong_statistical_physics}. The specific heat $C_V(T,N) = k_\text{B} \partial_T\left(T^2 \partial_T \ln Z(N,T) \right)$, utilizing Eq.\eqref{eq:appb-2}, equals:
    \begin{align}
        C_V(T,N) &= Nk_\text{B} \partial_T\left(T^2 \partial_T \left( - \gamma \frac{ \sqrt{2} \hbar g}{ck_\text{B} T} +\ln \frac{ck_\text{B} T}{\sqrt{2} \hbar g}  \right) \right) \nonumber \\
        &=Nk_\text{B} \partial_T\left(T^2\left(\gamma \frac{ \sqrt{2} \hbar g}{ck_\text{B} T^2} + \frac{1}{T}   \right)\right) = Nk_\text{B}.
    \end{align}
    This result does not yield meaningful physical insights; therefore, a perturbative quantum treatment is presented in Sec.~\ref{sec:iv}.

\section{Perturbative approximation for the Rindler Hamiltonian} \label{C}
We begin by identifying the unperturbed Hamiltonian \( \hat{H}_0 \) and the perturbation \( \hat{H}' \):
\begin{align} \label{eq:rind-ham-1}
    \hat{H}_0 &= mc^2 + \frac{\hat{p}^2}{2m} + mg\hat{x},\\
    \hat{H}' &= \frac{g\hat{x}\hat{p}^2}{2mc^2} - \ii \gamma \frac{\hbar g \hat{p}}{mc^2}. \label{eq:rind-ham-2}
\end{align}
The unperturbed Hamiltonian \( \hat{H}_0 \) describes a non-relativistic particle in a linear gravitational potential \( mg\hat{x} \). Its eigenstates \( |n\rangle \) and eigenenergies \( E_n^{(0)} \) are expressed in terms of Airy functions \cite{}:
\begin{align}
    E_n^{(0)} &= mc^2 + \epsilon_n,\\
    \pbraket{x}{n} &= \psi_n(x) = \frac{1}{\sqrt{L}} \Ai{\frac{x}{L} - a_n}, \label{eq:appc-airy-f}
\end{align}
where the characteristic length $L = \left( \frac{\hbar^2}{2m^2g} \right)^{1/3}$, while the energy shift $\epsilon_n =  a_n mgL$ with $a_n$ being one of the zeros of the Airy function $\Ai{-a_n} = 0$.

For the perturbation expansion, it will be beneficial to calculate the projection of operators $\hat{x}\hat{p}^2$ and $\hat{p}$ on the Airy function basis. The first one equals:
\begin{align}
    \ppbraket{k}{\hat{x}\hat{p}^2}{n} &= 2m \ppbraket{k}{\hat{x}\big(\hat{H}_0 - mc^2 - mg \hat{x}\big)}{n}\nonumber \\
    &= 2m\epsilon_n \ppbraket{k}{\hat{x}}{n} - 2m^2g \ppbraket{k}{\hat{x}^2}{n}, \label{eq:xp2}
\end{align}
where we have substituted the operator $\hat{p}^2$ from the unperturbed Hamiltonian given by the Eq.\eqref{eq:rind-ham-1}. The second one is derived from the commutation relation between $\hat{H}_0$ and $\hat{x}$, that equals:
\begin{align} \label{eq:com_H_x}
   [ \hat{H}_0, \hat{x} ] = \frac{1}{2m} [\hat{p}^2,\hat{x}] =  - \ii\frac{ \hbar \hat{p} }{ m }.
\end{align}
Evaluating the projection of the commutator Eq.\eqref{eq:com_H_x} onto the Airy basis, we obtain after algebra:
\begin{align} \label{eq:p}
    \ppbraket{k}{p}{n} = -\ii\frac{m}{\hbar} ( \epsilon_n - \epsilon_k ) \ppbraket{k}{x}{n}.
\end{align}
With Eqs.\eqref{eq:xp2}-\eqref{eq:p} at hand, the first-order energy correction due to the perturbation $\hat{H}'$ is given by the standard expression: 
\begin{align} 
    E_n^{(1)} &= \ppbraket{n}{\hat{H}'}{n}= \frac{g}{2mc^2} \ppbraket{n}{\hat{x} \hat{p}^2}{n} - \ii \gamma \frac{\hbar g}{mc^2} \ppbraket{n}{\hat{p}}{n} \nonumber \\
    &= \frac{g}{c^2} \left(\epsilon_n \ppbraket{n}{\hat{x}}{n} - mg \ppbraket{n}{\hat{x}^2}{n} \right)  \nonumber \\
    &= \frac{mg^2L^2}{c^2} \left(a_n \ppbraket{n}{\hat{\xi}}{n} - \ppbraket{n}{\hat{\xi}^2}{n} \right),
\end{align}
with non-dimensional operator $\hat{\xi} = \frac{\hat{x}}{L}$. Analogically, the second-order correction takes the form: 
\begin{align} 
    E_n^{(2)} &= \sum_{k \neq n} \frac{|\ppbraket{k}{\hat{H}'}{n}|^2}{\epsilon_n-\epsilon_k} = \sum_{k \neq n} \frac{\left| \frac{g}{2mc^2} \ppbraket{k}{\hat{x}\hat{p}^2}{n} - \ii \gamma \frac{\hbar g}{mc^2} \ppbraket{k}{\hat{p}}{n} \right|^2}{\epsilon_n-\epsilon_k} \nonumber \\
    &= \frac{g^2}{c^4} \sum_{k \neq n} \frac{\left| \left(\epsilon_n(1-\gamma) + \gamma\epsilon_k\right) \ppbraket{k}{\hat{x}}{n} - mg \ppbraket{k}{\hat{x}^2}{n}\right|^2}{\epsilon_n-\epsilon_k} \nonumber \\
    &=\frac{mg^3L^3}{c^4} \sum_{k \neq n} \frac{\left| \left(a_n(1-\gamma) + \gamma a_k\right) \ppbraket{k}{\hat{\xi}}{n} - \ppbraket{k}{\hat{\xi}^2}{n}\right|^2}{a_n-a_k}
\end{align}
The matrix elements $\ppbraket{k}{\hat{\xi}}{n}$ and $\ppbraket{k}{\hat{\xi}^2}{n}$ can be evaluated as follows. First, we write explicitly, using Eq.~\eqref{eq:appc-airy-f}:
\begin{align} \label{eq:appc-0}
    \ppbraket{k}{\hat{\xi}^j}{n} = \int \limits_{-\infty}^{+\infty} \dd{\xi} \, \xi^j \, \mathrm{Ai}\left[\xi - a_k\right] \mathrm{Ai}\left[\xi - a_n\right].
\end{align}
Then, the general method \cite{vallee2010airy, bieniek1977uniform} consists of using the integral representation of the Airy functions:
\begin{equation} \label{eq:appc-1}
\mathrm{Ai}\left[\xi - a_k\right] = \frac{1}{2\pi} \int_{-\infty}^{+\infty} \dd{t} \ee^{\ii\left(\frac{t^3}{3} + (\xi - a_k)\, t\right)}.
\end{equation}
With Eq.~\eqref{eq:appc-1} at hand, the Eq.~\eqref{eq:appc-0} becomes:
\begin{align}
\ppbraket{k}{\hat{\xi}^j}{n} = \int \limits_{-\infty}^{+\infty} \int \limits_{-\infty}^{+\infty} &\frac{\dd{t}\dd{t'}}{4\pi^2} 
\ee^{ \ii\left( \frac{t^3+t'^3}{3} - a_k t - a_n t' \right) } \int \limits_{-\infty}^{+\infty} \dd{\xi}  \xi^j \ee^{\ii\xi(t + t')},
\end{align}
and using the $j$th derivative of the Dirac delta function:
\begin{align}
\delta^{(j)}(t + t') &= \frac{(-\ii)^j}{2\pi} \int \limits_{-\infty}^{+\infty} \dd \xi \, \xi^j \ee^{\ii\xi(t + t')},
\end{align}
we finally find, after integration by parts, the relation:
\begin{equation}
\ppbraket{k}{\hat{\xi}^j}{n} = \int \limits_{-\infty}^{+\infty} \frac{\dd{t}}{2\pi}
\ee^{\ii\left(\frac{t^3}{3} - a_k t \right)} 
\dv[j]{}{t} \left( \ii^{-j} \ee^{-\ii\left( \frac{t^3}{3} - a_n t \right)} \right). \label{eq:appc-2}
\end{equation}
Explicitly, from Eqs.~\eqref{eq:appc-0}–\eqref{eq:appc-2}, we obtain:
\begin{align}
    \ppbraket{k}{\hat{\xi}}{n} &= -\frac{1}{2\pi} \int_{-\infty}^{+\infty} \dd t\, \ee^{\ii\left(a_n - a_k \right)t} \left(t^2 - a_n\right) \nonumber \\
    &= \delta''\left(a_n - a_k \right) + a_n \delta\left(a_n - a_k \right) \nonumber \\
    &\approx a_n \delta_{n}^{k} + \frac{\delta_{n}^{k+1} - 2\delta_{n}^{k} + \delta_{n}^{k-1}}{(a_{n+1}-a_{n})^2},
\end{align}
and
\begin{align}
    \ppbraket{k}{\hat{\xi}^2}{n} &= \frac{1}{2\pi} \int_{-\infty}^{+\infty} \dd t\, 
\ee^{\ii \left(a_n - a_k \right) t} \left( (t^2-a_n)^2 - 2\ii t \right) \nonumber \\
&= \delta^{(4)} \left(a_n - a_k \right) - 2a_n \delta''\left(a_n - a_k \right) \nonumber \\
& \quad + a_n^2 \delta\left(a_n - a_k \right) + 2 \delta'\left(a_n - a_k \right) \nonumber \\
&\approx a_n^2 \delta_{n}^{k} + \frac{\delta_{n}^{k+2} - 4\delta_{n}^{k+1} + 6\delta_{n}^{k}  - 4\delta_{n}^{k-1} + \delta_{n}^{k-2}}{(a_{n+1}-a_{n})^4} \nonumber \\
&\quad - 2 a_n \frac{\delta_{n}^{k+1} - 2\delta_{n}^{k} + \delta_{n}^{k-1}}{(a_{n+1}-a_{n})^2} + 2 \frac{\delta_{n}^{k+1} - \delta_{n}^{k-1}}{a_{n+1}-a_{n-1}},
\end{align}
where we have approximated the derivatives of the Dirac delta function using discrete Kronecker delta operators. 

This method was used in our numerical analysis: the discrete Kronecker delta approximations were inserted into the expressions for the matrix elements that determine the perturbative energy corrections, which were then summed to compute the full energy spectrum including ordering-dependent contributions. The resulting energies were subsequently used to evaluate the canonical partition function and to calculate the specific heat, allowing us to quantify the impact of higher-order corrections and the ordering parameters on thermodynamic properties.

\end{document}